\documentclass{article}
\usepackage{amsmath,amssymb}
\usepackage{epsfig}
\usepackage{array}
\usepackage[a4paper,top=3cm,bottom=2cm,left=3cm,right=3cm,marginparwidth=1.75cm]{geometry}

\title{Magnetic  diffusion and interaction effects on \\ Ultrahigh Energy Cosmic Rays: protons and nuclei}
\author{Juan Manuel Gonz\'alez, Silvia Mollerach and Esteban Roulet\\
Centro At\'omico Bariloche, Comisi\'on Nacional de Energ\'\i a At\'omica\\
Consejo Nacional de Investigaciones Cient\'\i ficas y T\'ecnicas (CONICET)\\
Av. Bustillo 9500, R8402AGP, Bariloche, Argentina}
\date{}

\begin{document}
\maketitle
\begin{abstract}
The flux of ultrahigh energy cosmic rays reaching the Earth is affected by the interactions with the cosmic radiation backgrounds as well as with the magnetic fields that are present along their trajectories.  
We combine the SimProp cosmic ray propagation code with a routine that allows to account for the average effects of a turbulent magnetic field on the direction of propagation of the particles.
We compute in this way the modification of the spectrum which is due to the magnetic horizon effect,  both for primary nuclei as well as for the secondary nuclei resulting from the photodisintegration of the primary ones.
We also provide analytic parameterizations of the attenuation effects, as a function of the magnetic field parameters and of the density of cosmic ray sources, 
which make it possible to obtain the expected spectra in the presence of the magnetic fields from the spectra that would be obtained in the absence of magnetic fields. The  discrete nature of the distribution of sources with finite density also affects the  spectrum of cosmic rays at the highest energies where the flux is suppressed due to the interactions with the radiation backgrounds, and parameterizations of these effects are obtained.
\end{abstract}

\section{INTRODUCTION}
\label{sec1}
Cosmic rays (CRs) are atomic nuclei that reach the Earth with energies extending up to beyond $10^{20}$\,eV, which are the highest particle energies that have ever been detected. Although their sources are still unknown,  CRs with energies above 1\,EeV (where 1~${\rm EeV}=10^{18}$\,eV) are expected to be of extragalactic origin. The main evidence for this comes from the fact that at energies of few EeV, for which the composition is relatively light \cite{augercomp}, the Galactic magnetic field is not strong enough to make CRs diffuse within the Galaxy, while on the other hand the arrival directions show no significant correlation with the distribution of Galactic matter \cite{augergc,tagc}. In addition, the observation of a dipolar distribution in the arrival directions of the CRs with energies above 8\,EeV, which points away from the Galactic center direction, also supports their extragalactic origin \cite{augerdip}. 

During their trip from the acceleration sites to the Earth, CRs can be deflected by the magnetic fields permeating the intergalactic space and they can also interact with the radiation in the cosmic microwave background (CMB) and the extragalactic background light (EBL), besides undergoing adiabatic losses  due to the cosmological expansion. CR protons lose energy mainly in interactions with the CMB, in particular by the photopion production at the highest energies as well as by pair production down to EeV energies. The main interactions affecting nuclei are the photodisintegration by interactions with the EBL and CMB, that changes the particle mass without changing their Lorentz factor, as well as the pair production that changes the Lorentz factor without changing the mass. These processes, as well as their effect on the spectrum and composition of the CRs arriving to the Earth, have been studied in great detail since the original works by Greisen, Zatsepin and Kuzmin \cite{gzk}, and there are some public numerical propagation codes which allow to compute them in full detail \cite{crpropa,SimProp}. 

The presence of intergalactic magnetic fields not only affects the arrival directions of the CRs but, if they are strong enough so as to prevent the low energy particles to reach the Earth, they can also modify their energy spectrum, a phenomenon known as the magnetic horizon effect \cite{lemoine,be07,bere08,gap08}. The magnitude of this effect depends on the strength of the magnetic field and on the typical distance between the sources. According to the propagation theorem \cite{propth}, the spectrum of the arriving particles would  not be affected by the presence of  magnetic fields as long as the distance between sources is much smaller than both the diffusion length and the attenuation length associated to the interactions. However, in the presence of sizeable magnetic fields and if CR sources are not very densely distributed, a suppression of the spectrum at low energies is expected.

This suppression has been quantified using a semianalytic approach in \cite{difu1} for the case of protons, with the results being also valid in the case of nuclei at the energies for which the interactions are negligible. In this paper we will extend the study to the case of primary nuclei in the regime in which interactions are relevant, analysing in detail the effects on the secondary particles produced by the photodisintegration during the propagation. To this scope we have extended the SimProp v2r4 code \cite{SimProp} so as to  not only follow the energy, mass and charge of the particles as a function of time, but also to follow the direction of propagation of the CRs and the distance from their sources, in the presence of a turbulent intergalactic magnetic field. In the simulations with the SimProp code we have used for definiteness  the photonuclear cross sections from Puget, Stecker and Bredekamp \cite{psb} and the extragalactic background light model from Stecker, Malkan and Scully \cite{EBLmodel}. We also study how a discrete source distribution affects the high energy cutoff due to the attenuation during propagation, as well as the possible recovery of the CR flux that may appear at the highest energies if the source spectra were to extend to extremely high energies.

\section{DIFFUSIVE CR PROPAGATION IN A TURBULENT MAGNETIC FIELD}

There are few observational constraints on the extragalactic magnetic fields (for a review see \cite{han17}). Precise values of their amplitudes are not known, and they  likely vary according to the region of space considered.
 In the  central regions of galaxy clusters the measured amplitudes range from a few up to tens of $\mu$G \cite{fe12}. This suggests that significant large-scale magnetic fields should also be present in cosmic structure filaments and sheets, while smaller strengths are expected in the void regions, with typical bounds in unclustered regions being $B<1$ to 10~nG.
 Realistic estimates for the magnetic fields in the Local Supercluster region range around 1 to 100~nG for their root mean square (rms) strength, and the coherence length may range from 10\,kpc to 1\,Mpc (see e.g. \cite{vallee11,fe12,enzo17}).  Note that the Galactic magnetic field, with typical strength of few $\mu$G, may affect the CR arrival directions, mainly through its regular component, but it is expected to have a subdominant effect on the CR spectrum due to its much smaller spatial extent, and it will hence be ignored here. 

Since the effects of the magnetic horizon become significant when the CRs are only able to reach the Earth from the closest sources, it is the magnetic field within the Local Supercluster which is most relevant, and thus we will not include larger scale inhomogeneities from filaments and voids.  We will then consider for simplicity the propagation of CRs in an homogeneous and isotropic turbulent extragalactic magnetic field. This can  be described by the rms amplitude, $B_{\rm rms}$, and the  coherence length, $L_{\rm coh}$, which is the maximum distance between two points for which the magnetic field is still correlated. One can define a critical energy  corresponding to the energy at which the Larmor radius of a particle with charge $eZ$  equals the coherence length, leading to $E_{\rm c} = Z |e| B_{\rm rms} L_{\rm coh} = 0.9 Z (B_{\rm rms}/{\rm nG})(L_{\rm coh}/{\rm Mpc})\, \rm{EeV}$. This energy separates the regime of resonant diffusion present at energies lower than $E_{\rm c}$ from  that  at higher energies in which the deflections  after traversing a distance $L_{\rm coh}$ are small.
In the latter case the diffusion can occur only for travelled distances much longer than the coherence length. 

The diffusion length $l_D$ is related to the diffusion coefficient $D$ by $l_D = 3 D/c$. It represents the average distance after which a particle is deflected by an angle of about 1~rad, so that the propagation of CRs from sources much more distant than $l_D$ will be diffusive. The dependence of the diffusion coefficient on the energy and magnetic field parameters which is inferred by following numerically the trajectories of many charged particles deflected by the Lorentz force is well fitted by the relation \cite{hmr14}

\begin{equation}
  l_D(E) = L_{\rm coh} \left[ 4 \left(\frac{E}{E_{\rm c}}\right)^2 + a_i \frac{E}{E_{\rm c}} + a_l \left(\frac{E}{E_{\rm c}}\right)^{\delta} \right],
    \label{eq:Ld}
\end{equation}
where for the case of a turbulent field with a Kolmogorov spectrum one has that $\delta = 1/3$,   $a_i \simeq 0.9$ and $a_l \simeq  0.23$.

The average effect of the turbulent magnetic field on the propagation of charged particles can be accounted for by integrating the stochastic differential equation \cite{achterberg,hmr16}
\begin{equation}
{\rm d}n_i =-\frac{1}{l_D}n_i c\,{\rm d}t + \frac{1}{\sqrt{l_D}} P_{ij}\,{\rm  d}W_j,
\label{eq:dni}
\end{equation}
where $\hat n\equiv (n_1,n_2,n_3)$ denotes the direction of the CR velocity and $P_{ij} \equiv (\delta_{ij} - n_i n_j)$ is the projection tensor onto the plane orthogonal to  $\hat n$. Repeated indices are summed and (${\rm d}W_1,\,{\rm d} W_2,\, {\rm d}W_3$) are three Wiener processes such that $\langle{\rm d} W_i \rangle =0$ and  $\langle {\rm d}W_i\,{\rm d}W_j \rangle =c\,{\rm d}t\,\delta_{ij}$.
In this way, the direction and distance from the initial point can be followed as the particles propagate.
Note that by solving this stochastic equation we are not integrating the CR trajectories in a fixed magnetic field realisation but rather obtaining the average behaviour of the deflections for different possible turbulent magnetic field realisations, which is the kind of information that we want to study in this work.

\subsection{Proton flux from one source}

The flux of protons expected from a CR source lying at a distance $r_{\rm s}$ much larger than the diffusion length can be computed by solving the diffusion equation  in an expanding universe, and is given by  \cite{Bere06}

\begin{equation}
    J_{\rm s}(E) = \frac{c}{4 \pi} \int_0^{z_{\rm max}} {\rm d}z \left| \frac{{\rm d}t}{{\rm d}z}\right| Q[E_{\rm g}(E,z),z] \frac{\exp[-r^2_{\rm s}/(4\lambda^2)]}{(4 \pi \lambda^2)^{3/2}} \frac{{\rm d}E_{\rm g}}{{\rm d}E},
    \label{eq:Bere}
\end{equation}
where $z_{\rm max}$ is the redshift at which the source started to emit, $E_{\rm g}(E,z)$  is the original energy at redshift $z$ of a particle reaching Earth with energy $E$, $Q$ is the source's emissivity spectrum  and $\lambda$ is the Syrovatskii generalised variable that takes into account the effects of the magnetic field, defined as
 
 \begin{equation}
   \lambda^2(E,z) = c\int^{z}_0 {\rm d}z' \left| \frac{{\rm d}t}{{\rm d}z'}\right| (1+z') ^2  \frac{l_{D}(E_{\rm g},z')}{3}.
    \label{eq:syrovat}
\end{equation}
One also has that
\begin{equation}
    \left| \frac{{\rm d}t}{{\rm d}z}\right|=\frac{1}{H_0(1+z)\sqrt{(1+z)^3\Omega_{\rm m}+\Omega_\Lambda}},
\end{equation}
with the Hubble parameter being $H_0\simeq 70$ km\,s$^{-1}$
Mpc$^{-1}$, the present matter content $\Omega_{\rm m}\simeq 0.3$ and the vacuum energy contribution $\Omega_\Lambda\simeq 0.7$.

Besides this analytic approach, we can also obtain the expected flux performing numerical simulations and following  up to $z=0$ the trajectories of a large number of particles with the desired distribution of initial energies and source emissivity, and selecting those that at the final time are located at a distance $r_{\rm s}$ from the initial point. As the particles propagate, the energy evolution is followed using the SimProp code, while the direction and position are followed by integrating eq.~(\ref{eq:dni}).

\begin{figure}[t]
    \centering
    \includegraphics[width=0.7\textwidth]{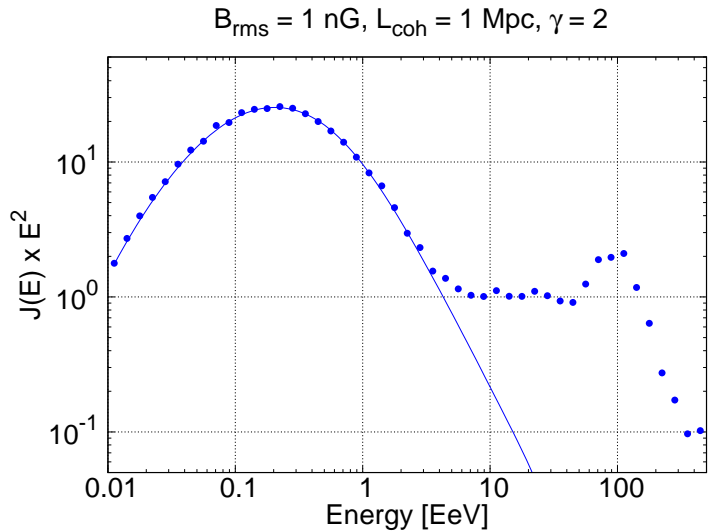}
    \caption{Flux coming from a proton source at 36\,Mpc emitting steadily since $z_{\rm max}=1$, adopting $B_{\rm rms}=1$~nG and $L_{\rm coh}=1$\,Mpc. The dots correspond to the numerical solution while the continuous line represents the solution to the diffusion equation  for the same conditions of propagation. Note the agreement between both solutions for energies where the diffusive regime holds (below 3\,EeV in this case), and the departure for energies where the quasirectilinear regime applies.}
    \label{fig:J_first_mio_Beren}
\end{figure}

Figure~\ref{fig:J_first_mio_Beren} shows the flux of protons expected from a source at a distance $r_{\rm s}=36$\,Mpc that emitted steadily with a spectrum $\propto E^{-2} $ since $z_{\rm max}=1$, immersed in a turbulent magnetic field with $B_{\rm rms} = 1$~nG and $L_{\rm coh} = 1$\,Mpc. The solid line is calculated using eq.~(\ref{eq:Bere}) while the dots are obtained from the integration of  eq.~(\ref{eq:dni}). For these values of the parameters, the transition between the diffusive and quasirectilinear propagation [$l_D(E) \simeq r_{\rm s}$] would take place  at an  energy of about 3\,EeV.
For lower energies the particles are in a diffusive propagation regime and the results of the stochastic propagation agrees with the solution of the diffusion equation. At higher energies, the propagation is quasirectilinear and the spectral shape is similar to the emitted one.  At even higher energies (above $\sim 100 \, \rm{EeV}$) the flux gets suppressed due to the interactions with the CMB. A pileup around 100\,EeV, just below the threshold for photopion production, is also clearly seen. 

\subsection{Flux from an ensemble of sources}

Let us now consider the flux expected from an ensemble of uniformly distributed equal luminosity sources with spatial density $n_{\rm s}$,  that for the commonly considered source populations typically takes values between $10^{-3}$ and $10^{-6}$\,Mpc$^{-3}$. The characteristic distance between sources is given by $d_{\rm s} \equiv n_{\rm s}^{-1/3}$, typically ranging from 10 to 100\,Mpc. 

After summation of the flux contribution given in eq.~(\ref{eq:Bere}) over all discrete sources, at distances $r_i$, one obtains that the total  flux is
\begin{equation}
J(E)\simeq \frac{R_H n_{\rm s}}{4\pi}\int_0^{z_{\rm max}}\frac{{\rm d}z}{(1+z)\sqrt{\Omega_m(1+z)^3+\Omega_\Lambda}}Q[E_{\rm g}(E,z),z]\frac{{\rm d}E_{\rm g}}{{\rm d}E} F,
\label{jza.eq}
\end{equation}
where $R_H= c/H_0 \simeq 4.3$~Gpc and
\begin{equation}
F\equiv \frac{1}{n_{\rm s}}\sum_i\frac{\exp(-r_i^2/4\lambda^2)}{(4\pi\lambda^2)^{3/2}}.
\label{f.eq}
\end{equation}

The generic implication of the diffusion effects, encoded in the factor $F$ which actually depends on the ratio $\lambda/d_{\rm s}$,  is to suppress the CR flux at low rigidities. This is because the particles take a much longer time than in the case of rectilinear propagation to arrive from the sources, and at low energies they may not be able to arrive to the observer even from the closest ones. However,  as long as the distance to the nearest sources  is smaller than the other relevant length scales (diffusion length and energy loss length), according to the  propagation theorem \cite{propth} one has that  the total CR flux will be the same as that obtained for a continuous distribution of sources and ignoring the magnetic field effects. This means that even at energies for which far away sources do not contribute anymore, as long as the observer lies within the diffusion sphere  of the nearby sources the spectrum will be  unchanged, and it is only when the nearest sources get also suppressed that the overall spectrum gets modified. This also shows that it is mainly the magnetic field in the local neighbourhood which is actually relevant to compute the magnetic suppression effects.

The validity of the propagation theorem can be understood by summing over the sources in eq.~(\ref{f.eq}),   and in the limit of small source separations replace the sum as $\sum\to n_{\rm s} \int {\rm d}r\,4\pi r^2$, and use that
\begin{equation}
\int_0^\infty {\rm d}r\,4\pi r^2 \frac{\exp(-r^2/4\lambda^2)}{(4\pi\lambda^2)^{3/2}}=1.
\label{one}
\end{equation}
Thus, for a continuous distribution of equal luminosity sources the diffusion effects do not modify the total flux.

The magnetic horizon effect  can be described through the flux suppression factor \cite{difu1}
\begin{equation}
G(E/E_{\rm c})\equiv \frac{J_Z(E)}{J_Z(E)|_{d_{\rm s} \rightarrow 0}},
\end{equation}
which is the ratio between the actual flux of  the CRs with charge $Z$ arriving to the observer and the one that would be obtained in the case of a continuous source distribution, corresponding to setting $F=1$ in eq.~(\ref{jza.eq}). For the summation over discrete sources we will adopt in the following the distances $r_i$ as being the average distances to the $i$th nearest source obtained in the case of a uniform source density, that is given by $r_i=(3/4\pi)^{1/3} d_{\rm s} \Gamma(i+1/3)/(i-1)!$. 
 The suppression factor $G$ depends on  the average distance between sources, $d_{\rm s}$, and on the coherence length, $L_{\rm coh}$, through the combination 
\begin{equation}
X_{\rm s}\equiv \frac{d_{\rm s}}{\sqrt{R_H L_{\rm coh}}}\simeq \frac{d_{\rm s}}{65\ \rm Mpc}\sqrt{\frac{\rm Mpc}{L_{\rm coh}}}.
\label{xs.eq}
\end{equation}
 
\begin{figure}[h]
    \centering
    \includegraphics[width=0.5\textwidth]{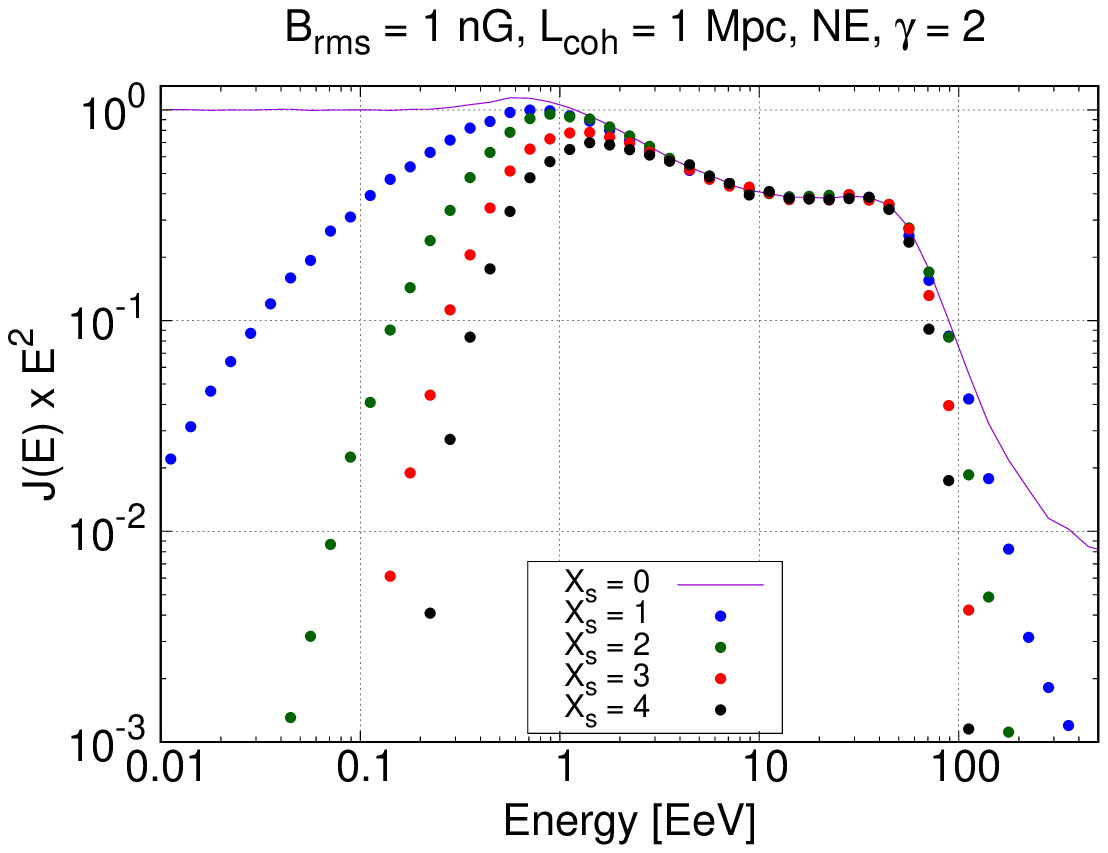}\includegraphics[width=0.5\textwidth]{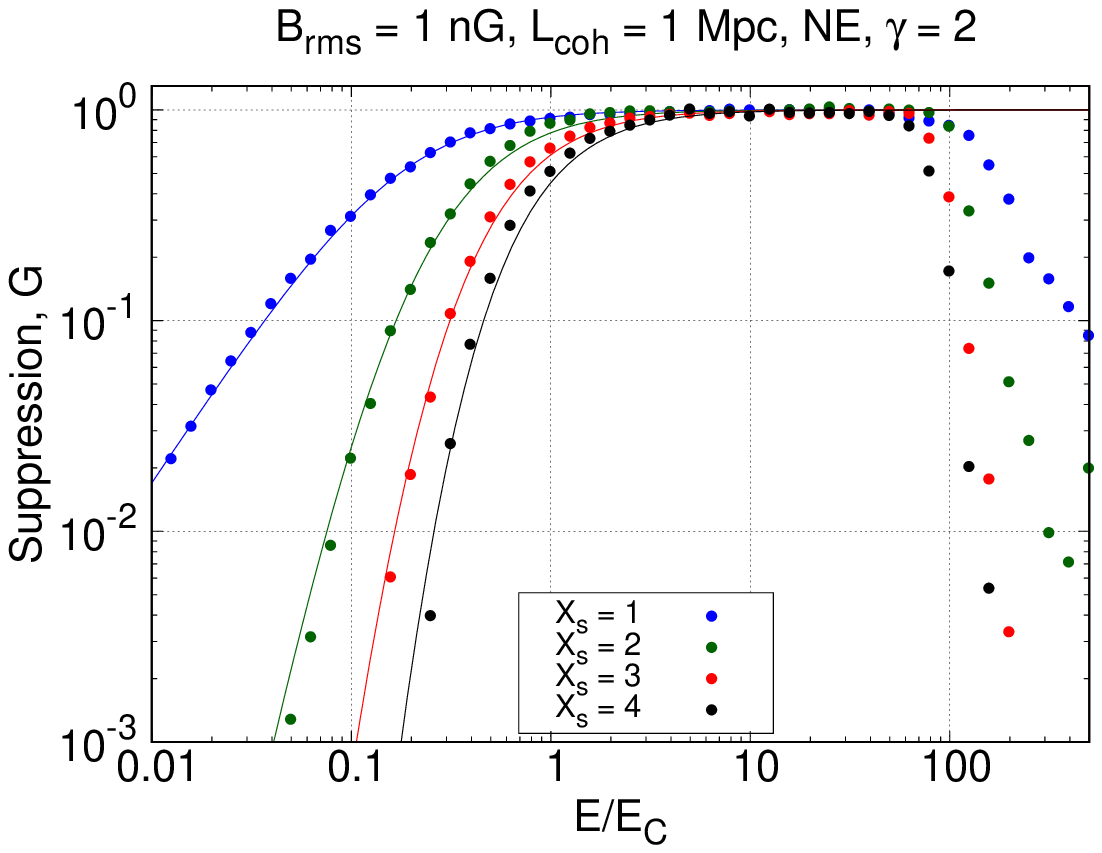}
    \caption{Left panel: Flux from primary protons for different  densities of sources (coloured points) and for a continuous distribution (purple line), adopting nonevolving sources with a spectral index $\gamma=2$. Right panel: Associated suppression factors due to the magnetic horizon (dots) and corresponding fits (continuous lines). At the highest energies also  the attenuation effects are apparent.}
    \label{fig:J_H_Fe}
\end{figure}

The left panel in Figure~\ref{fig:J_H_Fe} shows with dots the flux of protons  coming from an ensemble of sources having different average distances between them, and with a solid line that from a continuous distribution of sources (see e.g. \cite{be05}). In this and the following plots, unless specified otherwise,  we adopt as reference a uniform extragalactic magnetic field\footnote{The coherence length is assumed to be stretched by the expansion, so that $L_{\rm coh}(z)=L_{\rm coh}(0)/(1+z)$, while MHD considerations suggest \cite{be07} that $B(z)=(1+z)B(0)$.} with $B_{\rm rms}=1$~nG and $L_{\rm coh}=1$\,Mpc, considering sources with a spectral index $\gamma=2$ and a maximum energy of $10^5$\,EeV. We also consider that the sources are emitting with a constant luminosity since $z_{\rm max} = 1$ [no evolution case, (NE)].
At low energies, the departure between the continuous ($X_{\rm s} = 0$) and finite ($X_{\rm s} = 1, 2, 3, 4$) density cases is clearly seen. 
In the right panel the corresponding magnetic  suppression factor $G$ is plotted. This factor can be fitted using the expression \cite{mr20}
 \begin{equation}
   G(x) = \exp \left[ - \left(\frac{a X_{\rm s}}{x + b(x/a)^{\beta}} \right)^{\alpha} \right], 
    \label{eq:G_fit}
\end{equation}
where $x \equiv E/E_{\rm c}$. 
For the general case of a spectrum $\propto E^{-\gamma}$, but still considering nonevolving sources, good fits to the results are obtained with the parameters $a=0.206 + 0.026 \gamma$, $b=0.146 + 0.004 \gamma$, $\alpha=1.83 - 0.08 \gamma$ and  $\beta=0.13$ \cite{mr20}. The fitting functions for the particular case with $\gamma=2$ that was considered in Figure~\ref{fig:J_H_Fe} are shown in the right panel with continuous lines, for the different values of $X_{\rm s}$ displayed.

\begin{figure}[t]
\centering
\includegraphics[scale=0.6,angle=0]{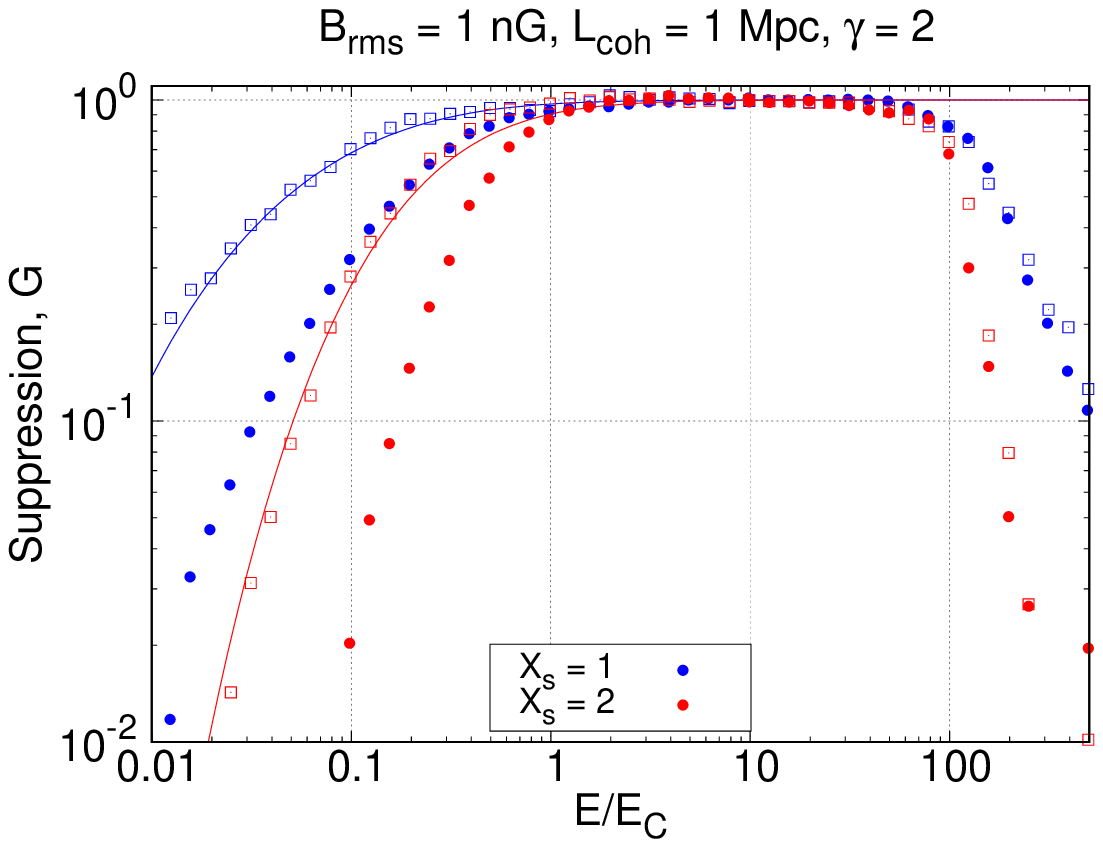}\includegraphics[scale=0.6,angle=0]{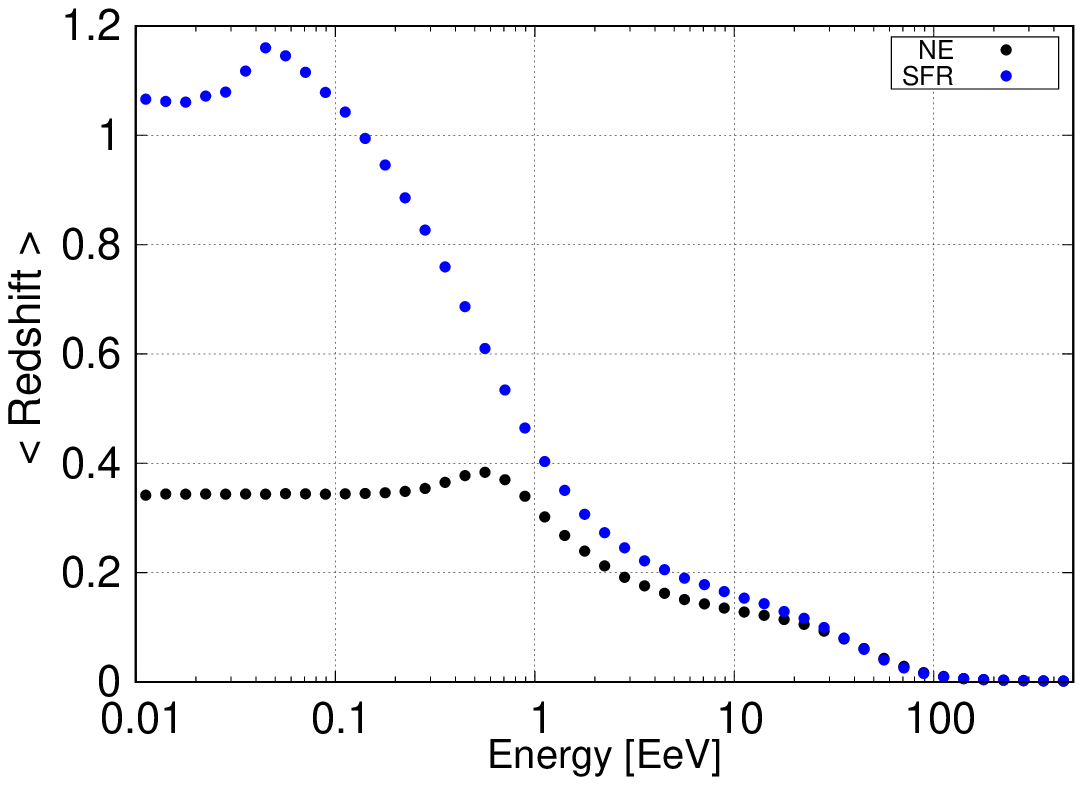}
\caption{Left panel: Magnetic suppression for protons in the NE (filled in points) and SFR (empty squares) scenarios for two different intersource separations. Continuous lines are the corresponding fits for the SFR case (see text). Right panel: Mean redshift at which protons originate as a function of the  observed energies, for the NE and SFR evolution scenarios.}
\label{fig:SFR_NES}
\end{figure}

The magnetic suppression actually depends on the evolution of the luminosity of the sources with redshift. As an example, we also consider the case of  sources emitting proportionally to the star formation rate (SFR), for which we adopt the parametrization from  \cite{Hopkins}, assuming that the source emissivity scales as $Q \propto (1+z)^{3.44}$ up to $z = 0.97$. For larger redshifts and up to $z = 4.48 $ it falls  as $(1+z)^{-0.26}$. This falloff becomes steeper for higher redshifts, but anyhow we simulate particles in this case up to a maximum redshift $z_{\rm max}=4$, since the contribution from higher redshifts is negligible.

The SFR scenario  leads to a milder magnetic suppression in the flux detected at Earth, as shown in the left panel of Figure~\ref{fig:SFR_NES}.
The magnetic suppression for the SFR case can also be parametrized with the same functional form given in eq.~(\ref{eq:G_fit}),  with parameters $a=0.135 + 0.04 \gamma$, $b=0.254 + 0.04 \gamma$, $\alpha=2.03 - 0.11 \gamma$ and $\beta=0.29$ \cite{mr20}, and the fits for $\gamma=2$ are displayed in the figure with solid lines.
For a given arrival energy, in the SFR scenario  the  particles come on average from a higher redshift than in the NE scenario, as shown in the right panel. The longer associated travel times mean that particles can reach the Earth on average from larger distances, and this is why the magnetic suppression is weaker.

In all the cases, a suppression of the flux at high energies due to the interactions with the radiation backgrounds is also apparent, and we will discuss it in more detail in Section~\ref{sec:Factor_H}.

\section{MAGNETIC DIFFUSION OF HEAVIER NUCLEI AND OF THEIR SECONDARIES}

The effect of the turbulent magnetic field on the flux expected from sources emitting nuclei can be computed in a similar way by following the energy, charge and mass of the primary particle, as well as of the secondary fragments produced, using the SimProp code to account for the interactions and following the direction of propagation and distance from the source by integrating the stochastic differential equation in eq.~(\ref{eq:dni}).  

\begin{figure}[t]
\centering
\includegraphics[scale=0.61,angle=0]{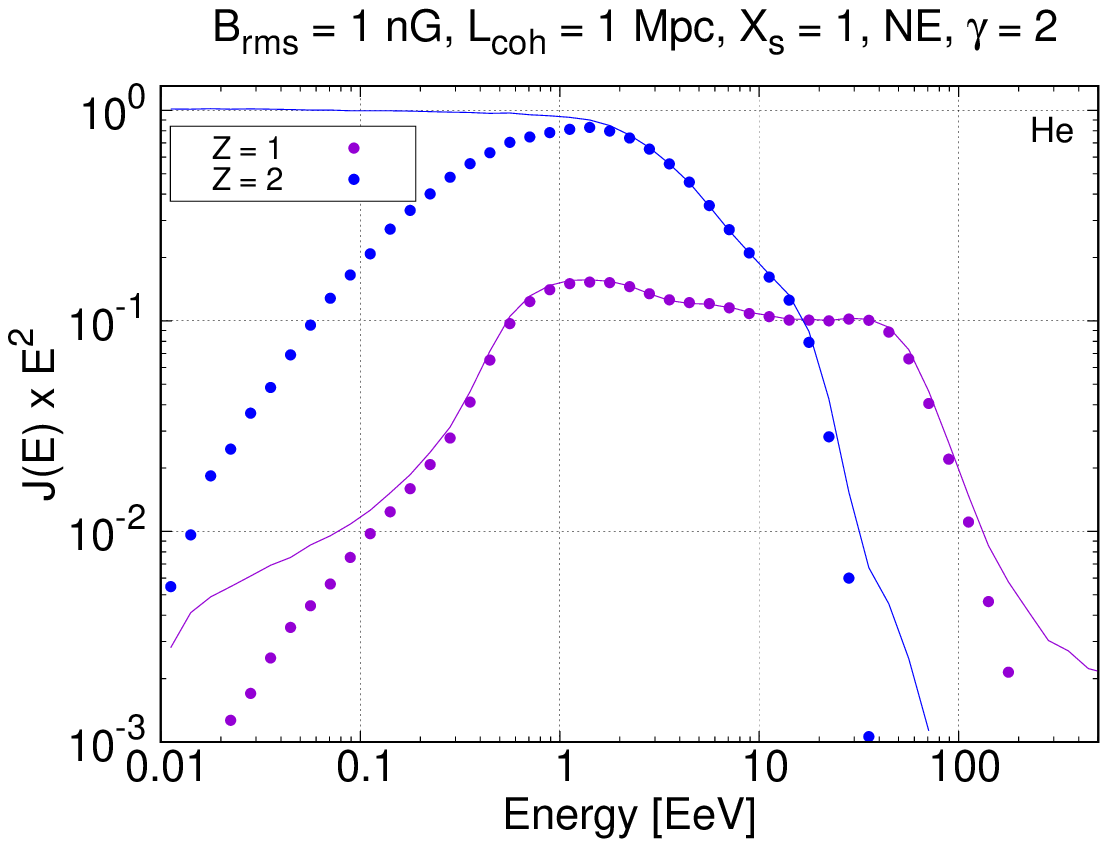}\includegraphics[scale=0.6,angle=0]{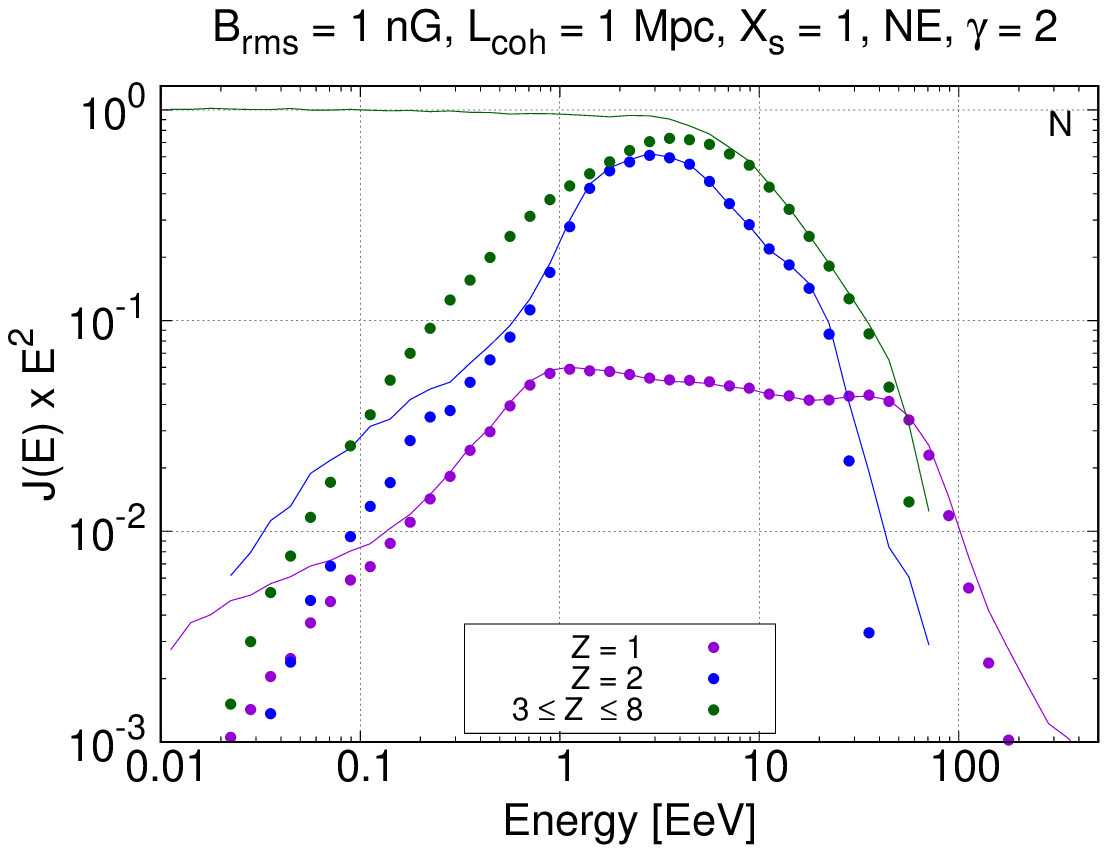}
\includegraphics[scale=0.61,angle=0]{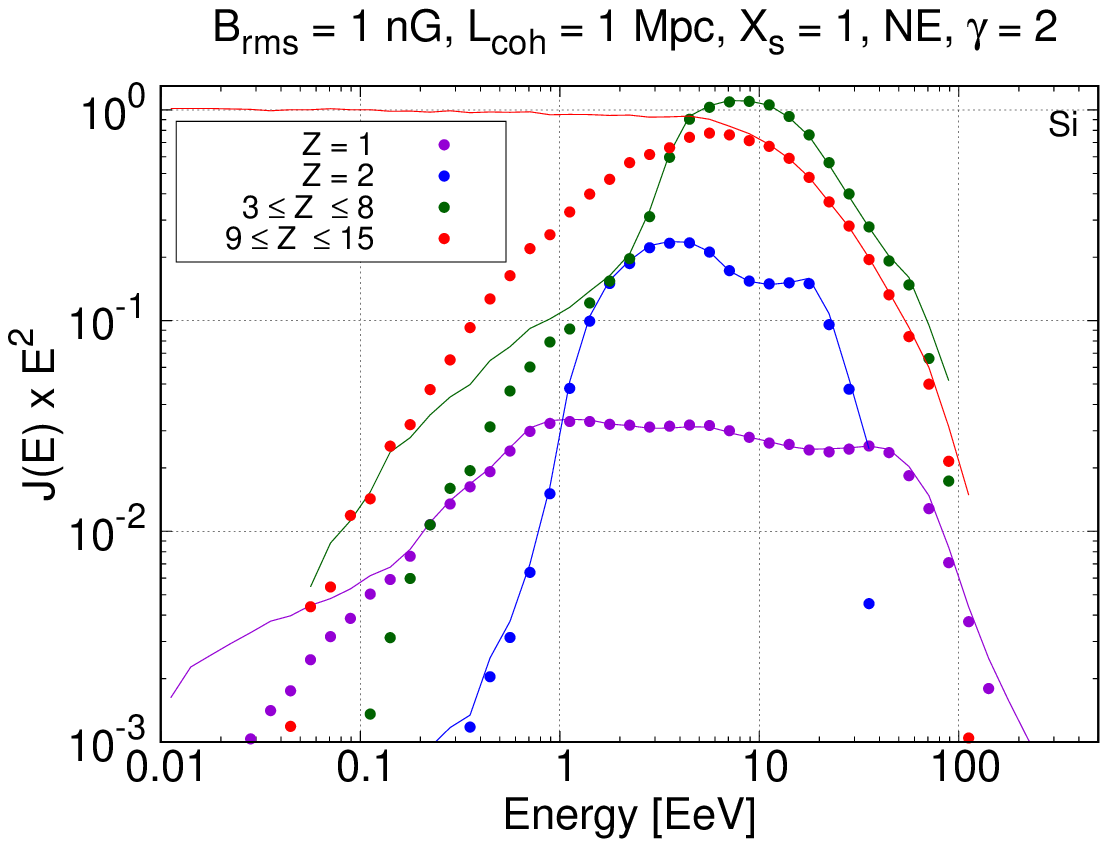}\includegraphics[scale=0.6,angle=0]{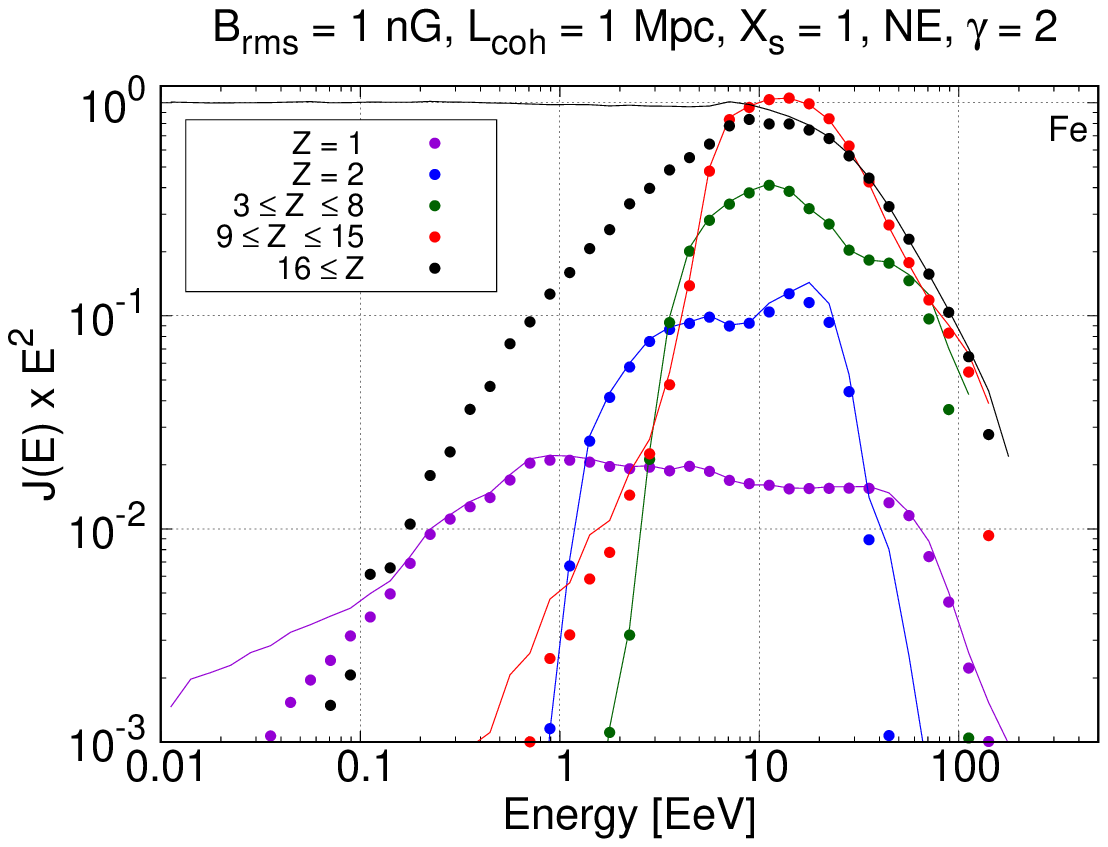}
\caption{Flux of primary nuclei and of their secondaries for the different cases of primary helium, nitrogen, silicon and iron, as indicated.}
\label{fig:J_sec}
\end{figure}

In order to study  different possible primaries, we followed the evolution for the illustrative cases of He, N, Si and Fe nuclei. We then collected the final nuclei at $z=0$ in four different groups according to their charge:  $Z=2$ (He), $ 3 \leqslant Z \leqslant 8$ (CNO group), $9 \leqslant Z \leqslant 15$ (Si group) and
$ 16 \leqslant Z \leqslant 26$ (Fe group), considering also separately the secondary nucleons. 
Figure~\ref{fig:J_sec}
shows with solid lines the expected flux of particles coming from a continuous distribution of sources  and with dots those from a discrete distribution with $X_{\rm s} = 1$, corresponding to a separation  $d_{\rm s}=65$\,Mpc  for the adopted value of $L_{\rm coh}=1$\,Mpc (and hence a density $\sim  4\times 10^{-6}$\,Mpc$^{-3}$).

\begin{figure}[h]
\centering
\includegraphics[scale=0.6,angle=0]{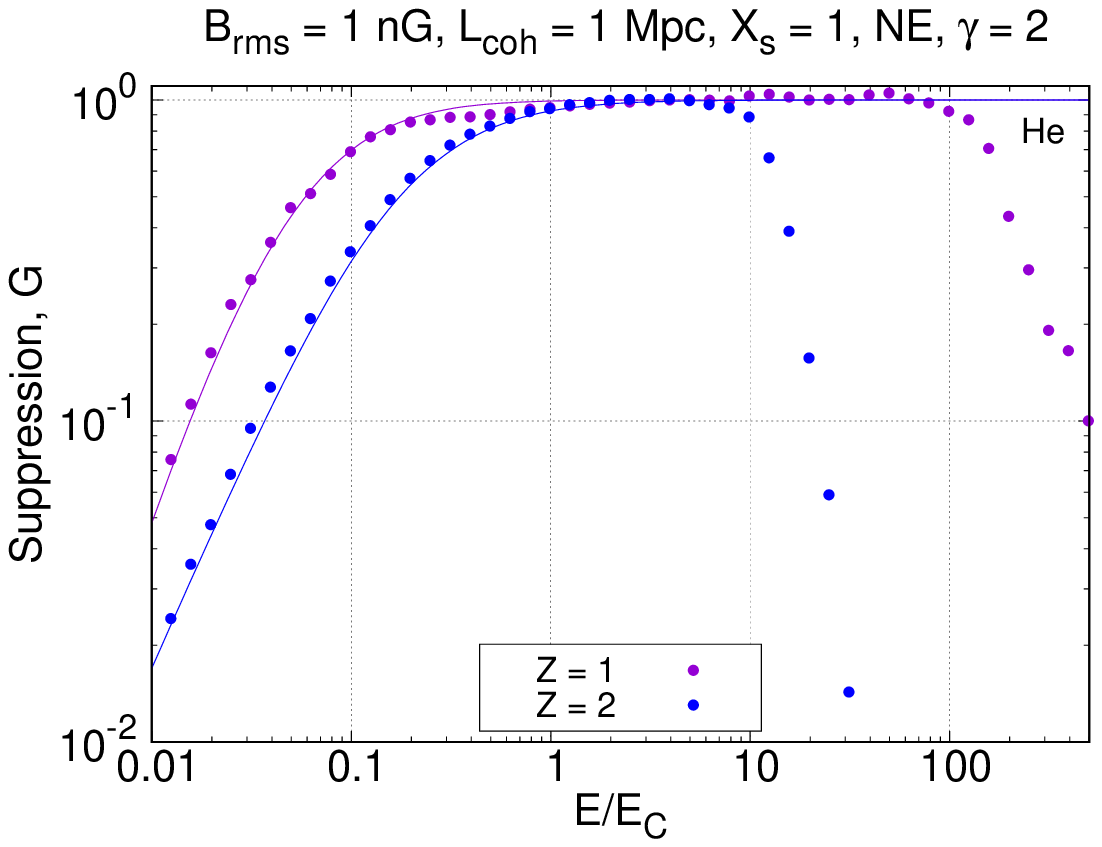}\includegraphics[scale=0.6,angle=0]{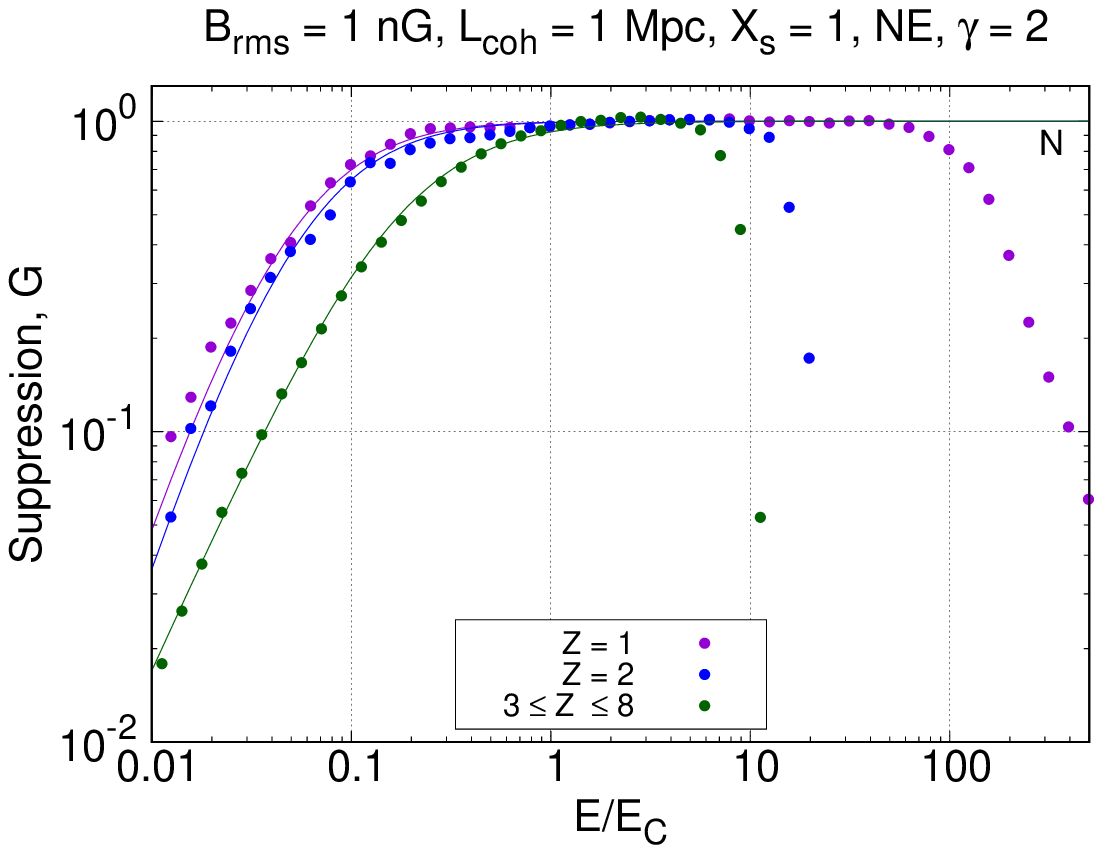}

\includegraphics[scale=0.6,angle=0]{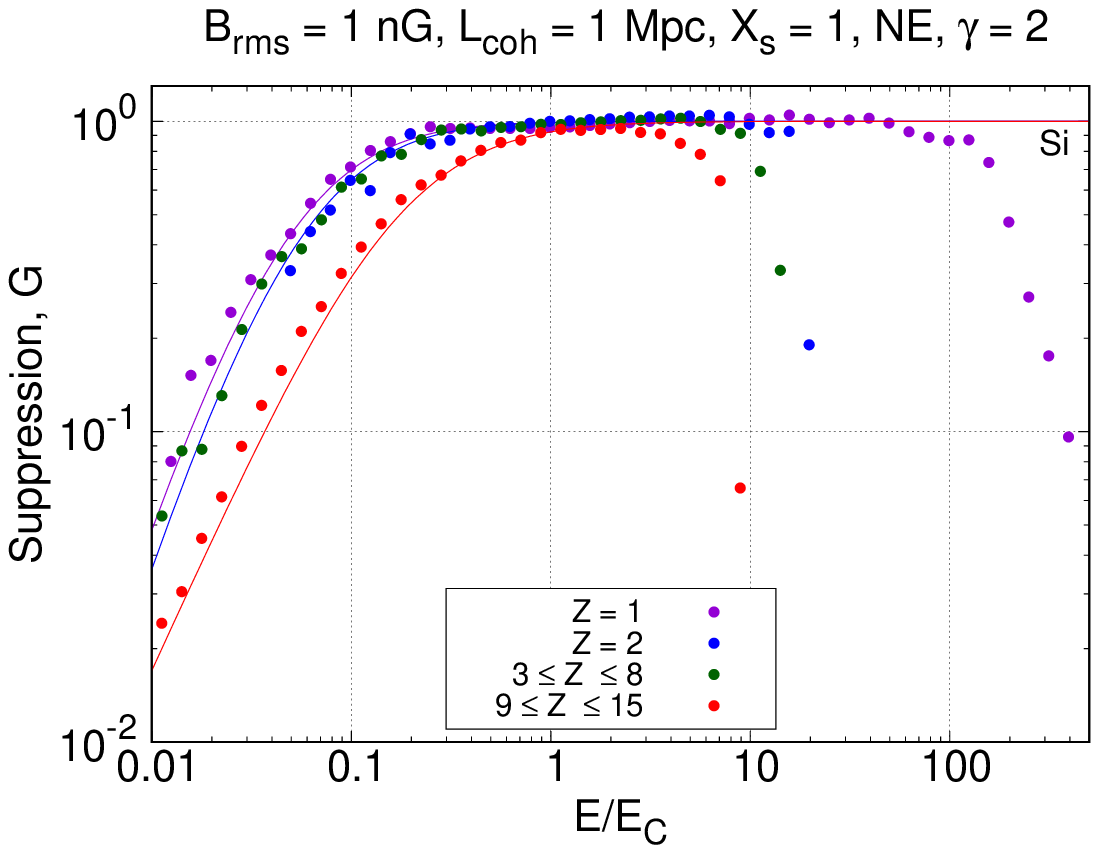}\includegraphics[scale=0.6,angle=0]{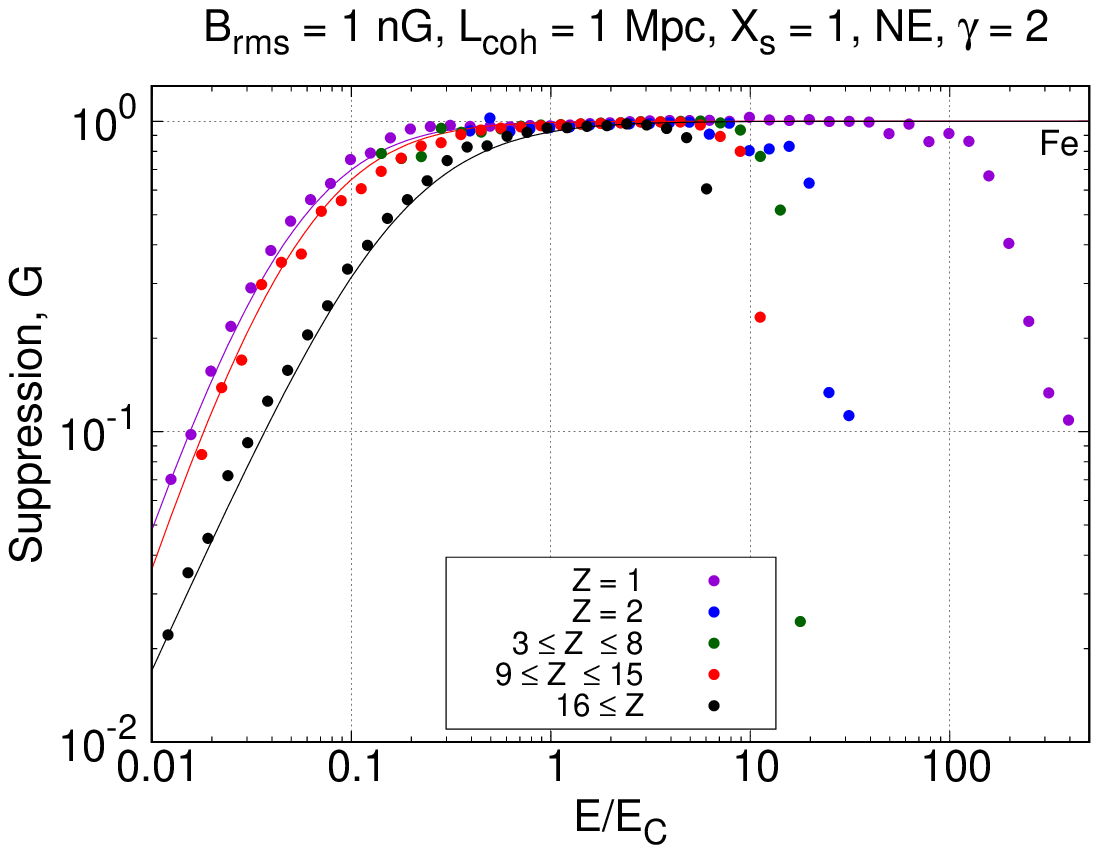}

\caption{Flux suppression factor for primary nuclei and their secondaries for the different cases of primary helium, nitrogen, silicon and iron, as indicated. Solid lines correspond to the fits to the magnetic suppression factor $G$ described in the text.}
\label{fig:G_sec}
\end{figure}

 In Figure~\ref{fig:G_sec} we show the ratio between the flux for the discrete source distribution and that for the continuous  one for the four primary masses considered and for the different groups of secondaries. The magnetic suppression effect appearing at low energies is clearly seen in all  cases. We see that for all the primaries considered the flux of the heavier leading fragments reaching the Earth with a mass in the same mass group as the original particle is more suppressed than that of lighter mass fragments. We  will now quantify the suppression for both sets of particles in more detail.

\subsection{Primary nuclei}

\begin{figure}[t]
    \centering
    \includegraphics[width=0.49\textwidth]{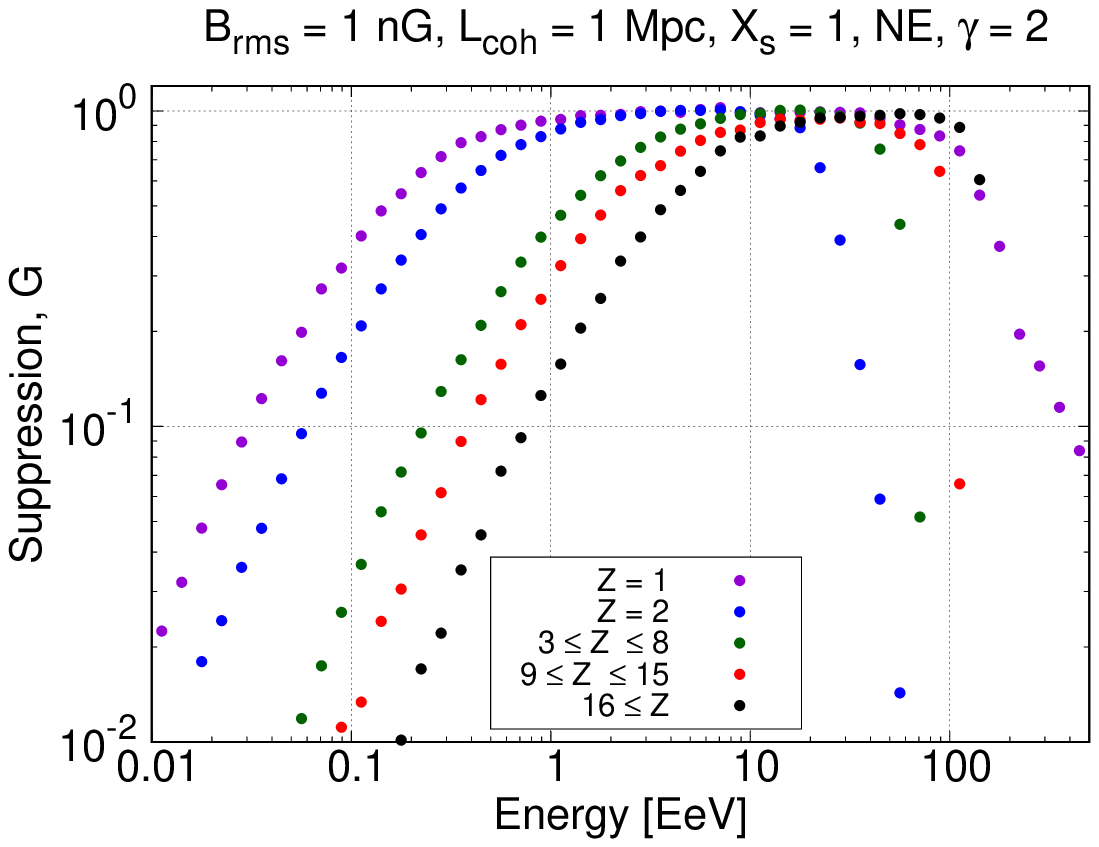}\includegraphics[width=0.49\textwidth]{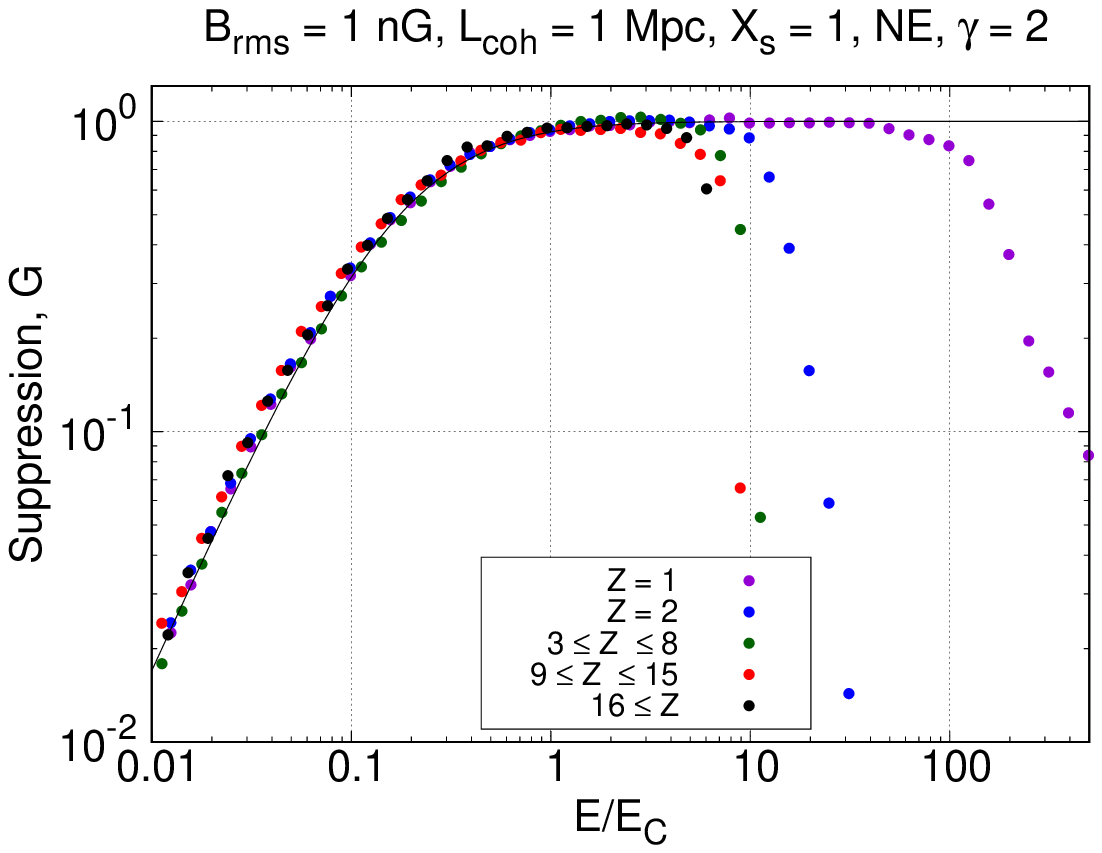}
    \caption{Left panel: Flux suppression  as a function of energy for the  higher mass leading fragments of each  primary nuclear species.  Right panel: Suppression factor for each mass group as a function of the energy divided by the critical energy. The solid curve represents the analytic fit to the magnetic suppression factor $G$ for the proton case.}
    \label{fig:G_5grupos}
\end{figure}

Let us first analyse the magnetic suppression of the flux at low energies for the particles arriving to the observer with a mass in the same mass group as the primary nuclei. We show in Figure~\ref{fig:G_5grupos} the corresponding 
suppression of the flux as a function of the energy and for  the different primary nuclei (left panel), for the same scenario considered before with $d_{\rm s}=65$\,Mpc.
At a given energy, the particles with larger charges, and hence  smaller rigidities, take more time to reach the Earth  and thus their flux gets more suppressed by the magnetic horizon effect. However, when plotted as a function of $E/E_{\rm c}$ (right panel)
the magnetic suppression for all the masses is instead  approximately the same, and it coincides with that discussed in the previous section for the case of protons.
This behaviour is the consequence of the fact that  the particles with similar rigidities have their trajectories similarly modified by the magnetic field.
Note that to compute the critical energy within each mass group, we consider for simplicity the  charge of the representative element for each group ($Z=7$ for the CNO group, $Z=14$ for the Si group or $Z=26$ for the Fe group).

\begin{figure}[t]
    \centering
    \includegraphics[width=0.49\textwidth]{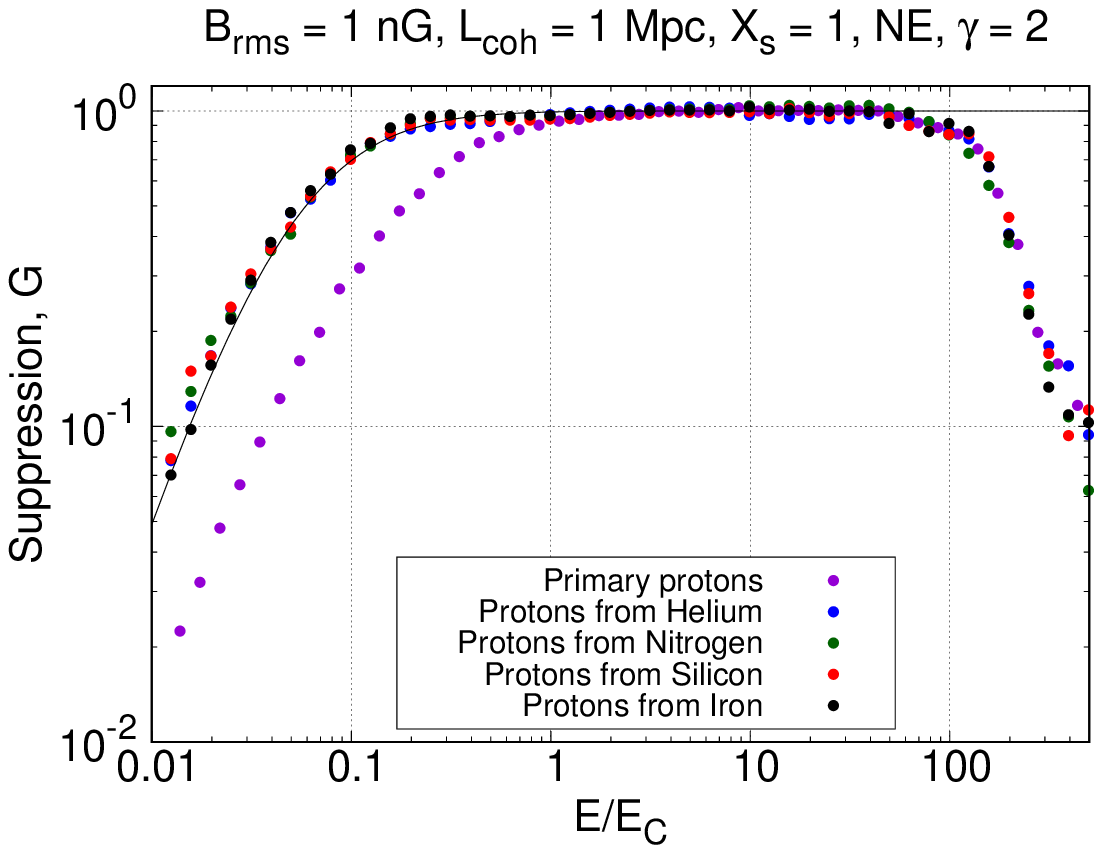}
    \includegraphics[width=0.49\textwidth]{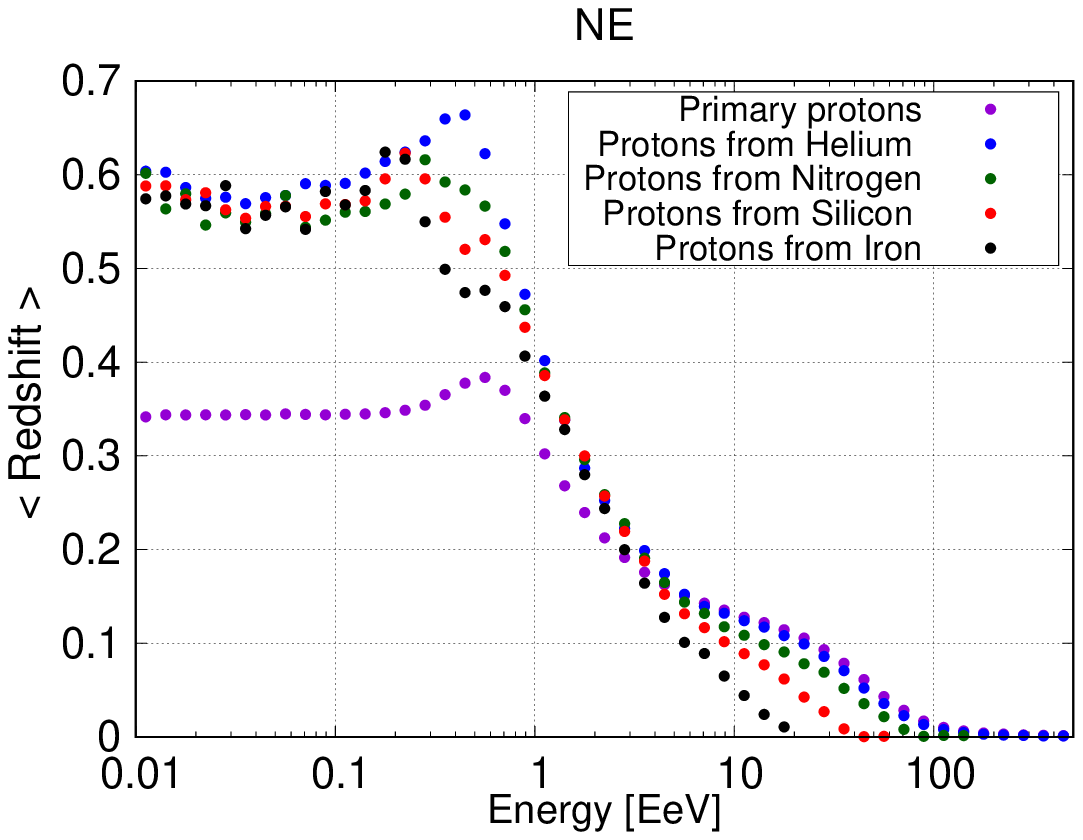}
    \caption{Left panel: Magnetic suppression for secondary protons from different primaries. Right panel: Average initial redshift of the primary nuclei that gave rise to the secondary protons, as a function of the energy of the secondary protons. In both panels the results for primary protons (purple dots) were included to facilitate the comparison.}
    \label{fig:Redshift_5_groups}
\end{figure}

\subsection{Secondary particles}

 As was shown in Figure~\ref{fig:J_sec}, the flux of secondary particles coming from the photodisintegration of heavier nuclei also has a magnetic suppression at low energies, and as one can see in Figure~\ref{fig:G_sec} their suppression  is milder than that of primary nuclei and  it is quite similar for the different secondary mass groups considered. 
In order to understand the differences in the resulting magnetic suppressions, we will first focus on the secondary protons produced from the propagation of different primaries and then extend the results to the heavier secondaries.

One can see from  the left panel of Figure~\ref{fig:Redshift_5_groups} that the suppression of the flux of the secondary protons from all the different primary nuclei is almost the same, having a similar shape  as that obtained for primary protons but being shifted to energies lower by a factor of about 2 to 3 for the parameters considered.  In the right panel we plot the mean original redshift of the primary nuclei that produced the secondary protons as a function of the final energy of the secondary proton. One can see that in the low energy range where the magnetic suppression of the flux takes place,  the primary nuclei have a mean redshift which is  larger than the one of the primary protons, and this holds for all the different nuclei considered. Longer travel times mean that more particles can reach the Earth at low energies, and this is the main reason leading to the milder suppression observed in the case of secondary protons. In addition to this effect, before the emission of the secondary proton the parent nucleus had a higher rigidity than the final proton (about twice as large), travelling then straighter and being able to arrive from sources farther away, becoming hence less affected by the magnetic effects than what a primary proton would  have been.

\begin{figure}[]
    \centering
    \includegraphics[width=0.49\textwidth]{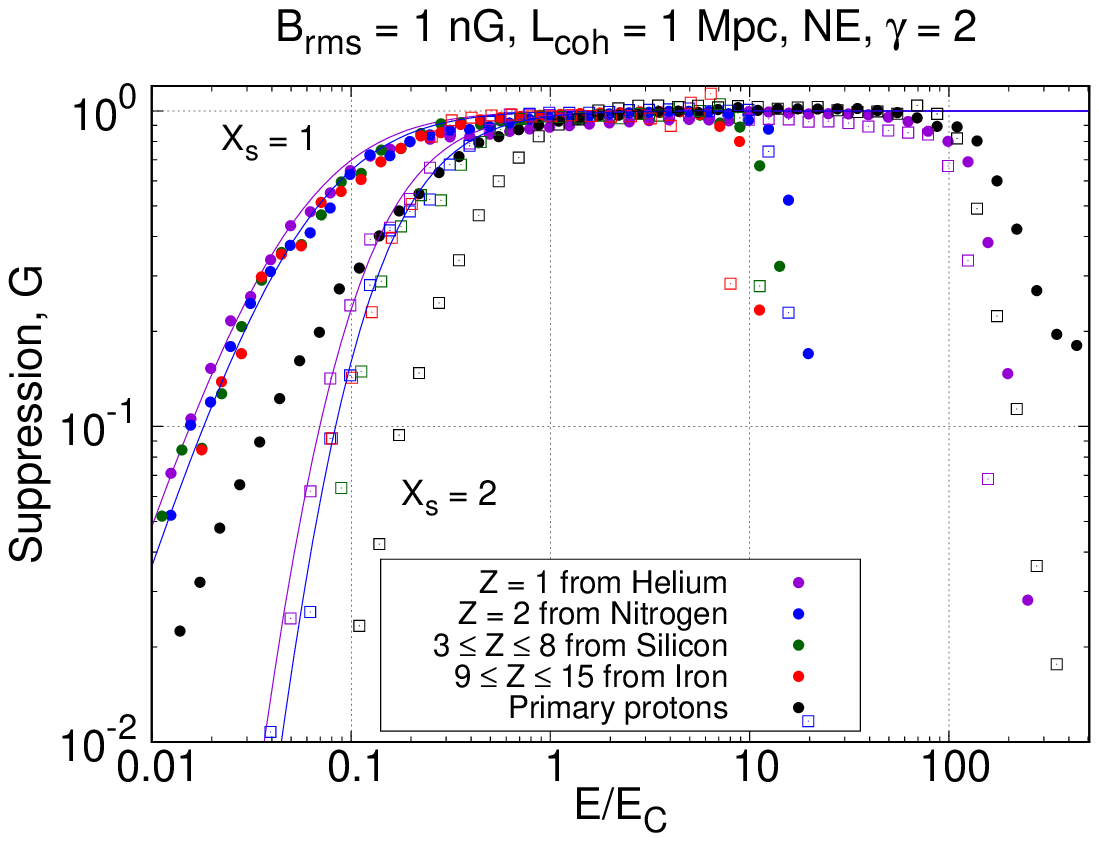}
    \includegraphics[width=0.49\textwidth]{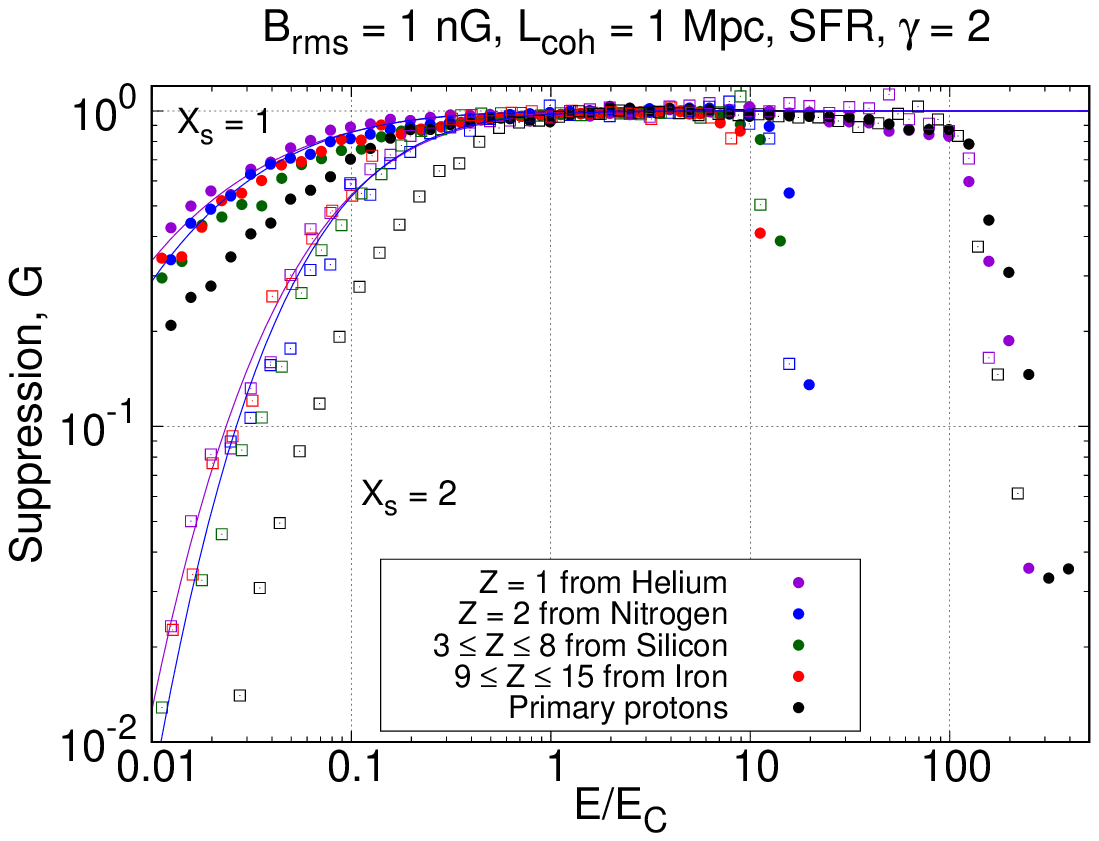}
    \caption{ Magnetic suppression for intermediate mass secondary nuclei and for primary protons for two values of the $X_{\rm s}$ parameter 
    ($X_{\rm s}=1$ with filled circles and $X_{\rm s}=2$ with empty squares) in the NE scenario (left panel) and for the SFR scenario (right panel). Purple and blue lines correspond to the fits to the magnetic suppression for secondary protons and intermediate secondary nuclei respectively with the parameters from Table \ref{table_1}.}
    \label{fig:intermediatesec}
\end{figure}

As it can be seen in Figure~\ref{fig:intermediatesec}, the magnetic suppressions of the fluxes of the remaining secondary nuclear fragments turn out to be close to that of the secondary protons, although they are slightly stronger given that the last effect mentioned above is not present. This figure shows the magnetic suppression for different secondary nuclei for two values of the density of sources, corresponding to $X_{\rm s}=1$ and $X_{\rm s}=2$. The left panel is for the case of no evolution of the sources, while the right panel shows the same results for the case of SFR evolution. The slight differences between the magnetic suppression factors from different secondary mass groups can  be understood from the differences in the average redshift of emission of the primaries that gave rise to the different lighter secondaries, which are  larger for the lighter secondaries than for the heavier ones given that more photodisintegrations are required to produce a lighter secondary.

Given that the suppression obtained for the lighter secondary nuclei are all quite similar,  we will just consider a common value for the suppression for all of them. To describe this suppression we also use the expression in  eq.~(\ref{eq:G_fit}), and the parameters of the different fits for all cases discussed in this work are collected in  Table~\ref{table_1}. These parameters  describe the spectral suppression at low energies for any value of the source density, as parameterized by the corresponding value of $X_{\rm s}$. 
%\bfseries
\begin{table}[ht]
\centering
  \renewcommand{\arraystretch}{1}
 \begin{tabular}{>{\centering}m{3cm} >{\centering}m{2.1cm} >{\centering}m{3cm} >{\centering}m{2cm} >{\centering\arraybackslash}m{2cm}}

%\hline
\hline\hline
  \multicolumn{5}{c}{NE} \\ 
\hline\hline
& $a$ &  $b$ & $\alpha$ & $\beta$\\ 
\hline

Primaries & $ 0.206 + 0.026\gamma  $ & $ 0.146 + 0.004\gamma  $& $ 1.83 - 0.08\gamma$ &   \\
\cline{1-4}
Secondary  \\* protons& $ 0.098 $ & $ 0.072 - 0.005\gamma  $& $ 2.02 $ & 0.129 \\
\cline{1-4}
Intermediate \\* secondary nuclei & $ 0.117 $ & $ 0.092 - 0.008\gamma $& $ 2.08 $ &\\
\hline\hline
  \multicolumn{5}{c}{SFR} \\ 
\hline\hline
& $a$ &  $b$ & $\alpha$ & $\beta$\\ 
\hline
Primaries & $ 0.135 + 0.040\gamma $ & $ 0.254 + 0.040\gamma $& $ 2.03 - 0.11\gamma $  & \\ 
\cline{1-4}
Secondary  \\* protons & $ 0.117 $ & $ 0.266 - 0.029\gamma $& $ 1.99 $  &  0.29 \\ 
\cline{1-4}
Intermediate \\* secondary nuclei & $ 0.103 $ & $ 0.242 - 0.040\gamma $& $ 2.01 $ & \\ 
%\hline
\end{tabular}
\caption{Parameters of the fit for the magnetic suppression factor $G$, using eq.~(\ref{eq:G_fit}),  for primary nuclei, secondary protons and intermediate mass secondary nuclei, for both the NE and SFR scenarios.}
\label{table_1}
\end{table}

\section{FLUX SUPPRESSION AT THE HIGHEST ENERGIES} \label{sec:Factor_H}

\begin{figure}[t]
    \centering
    \includegraphics[width=0.6\textwidth]{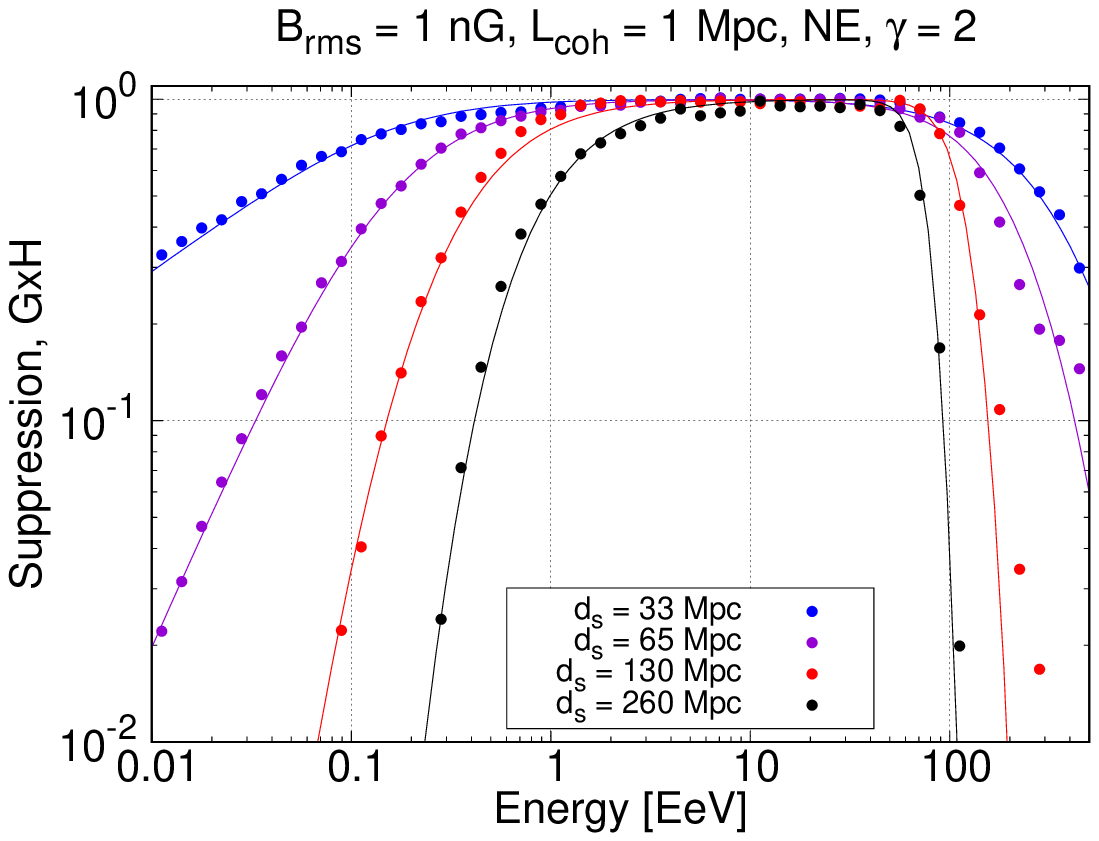}
    \caption{Suppression of the flux for protons for different distances between sources (points) and fitting functions G$\times$H.}
    \label{fig:GxH}
\end{figure}

 As can be seen in Figure~\ref{fig:G_5grupos}, when considering a finite source density one finds that besides the low energy suppression there is also a change in the suppression of the flux  at high energies with respect to the case of continuously distributed sources ($d_{\rm s} \rightarrow 0$) \cite{be07,bere08,gap08}. A similar effect would also appear for continuously distributed sources if one considers a minimum source distance \cite{taylor}. 
 Above 60\,EeV the attenuation length of protons due to the photopion production with the CMB rapidly shrinks, dropping well below 100\,Mpc for energies larger than 100\,EeV. Something analogous happens for heavier nuclei at energies larger than about $5Z$\,EeV, due to the photodisintegrations with the CMB photons.   Thus,  a sharp drop in the flux should appear for the energies at which the attenuation length becomes comparable or smaller than the distance to the closest sources. This high-energy suppression can be modelled using the following functional form
 \begin{equation}
    H(E) = \cosh^{-1} \left[ \left( \frac{E}{E_{\rm cut}}\right)^s\right],
\end{equation}
where $E_{\rm cut}$ represents a cutoff energy and the index $s$ controls the sharpness of the suppression. Figure~\ref{fig:GxH} shows the suppression of the proton flux for different average distances between the sources. A magnetic field was also included in the computations, thus also the magnetic suppression at low energies is present, but the deflections have no impact on the high-energy suppression since this one takes place in the regime of quasirectilinear propagation. We see that the total suppression is very well described by the product of the factors $G(E)$ and $H(E)$. 

\begin{figure}[t]
    \centering
    \includegraphics[width=0.5\textwidth]{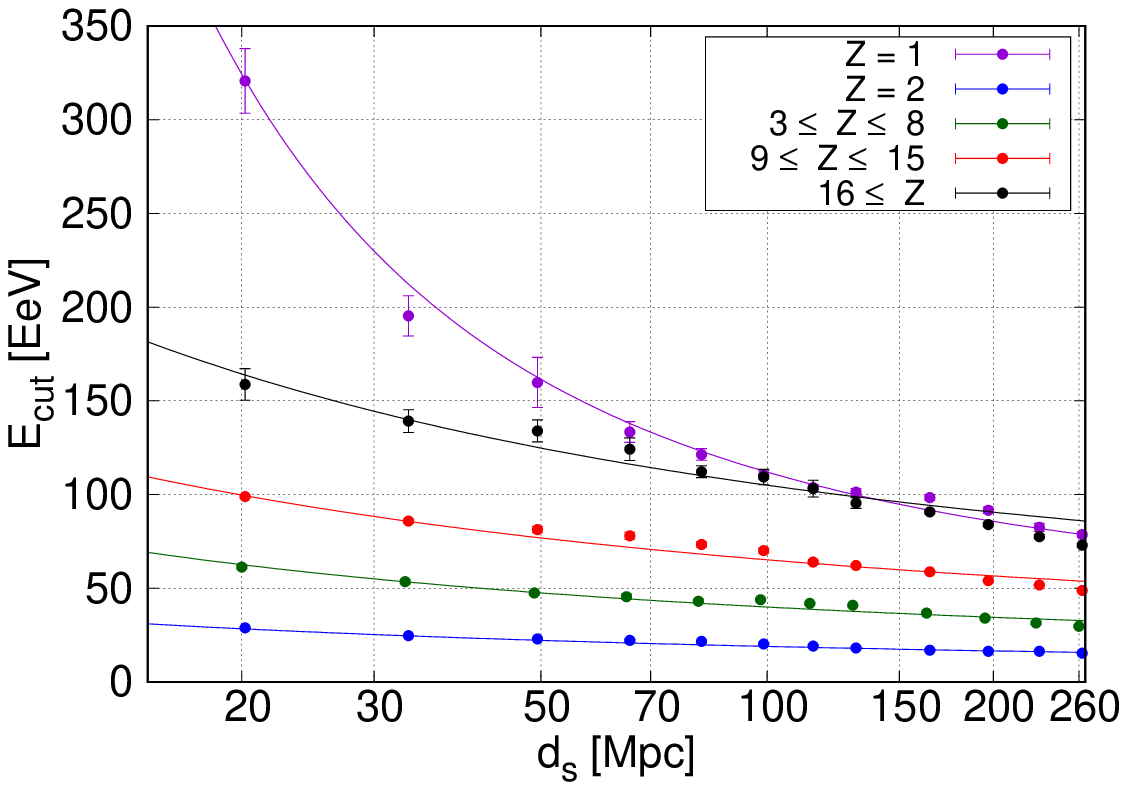}\includegraphics[width=0.5\textwidth]{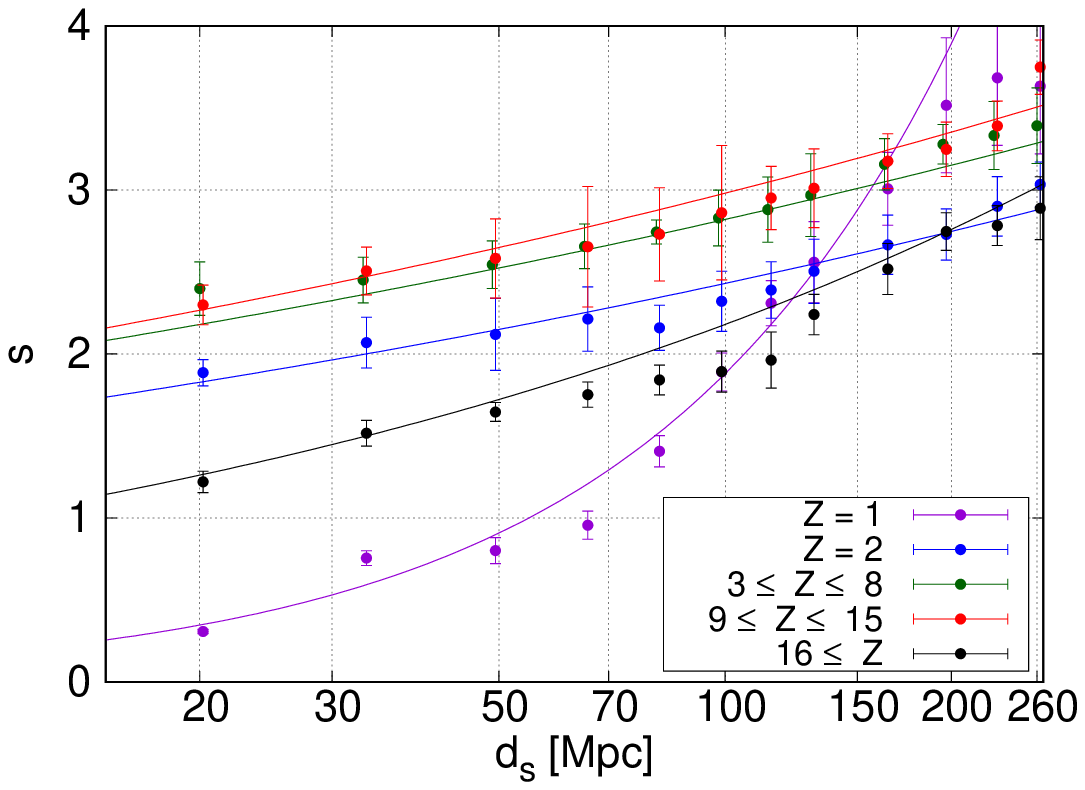}
    \caption{ Cutoff energies (left panel) and index $s$ (right panel) for the 5 mass groups, as a function of the characteristic distance between sources, for the case of a spectral index at the source $\gamma=2$.}
    \label{fig:E_cuts}
\end{figure}

The dependence of the cutoff energy and of the index $s$ on the mean source separation, for particles representative of the five mass groups considered, are shown  in Figure~\ref{fig:E_cuts} for the case of a spectral index at the source $\gamma=2$. As expected, the cutoff energy  decreases  for increasing intersource distances, since this leads to stronger attenuations caused by the interactions with the radiation backgrounds. We also find that for increasing intersource distances the suppression of the spectrum  at  high energies becomes steeper, what corresponds to larger values of the index $s$. 
Solid lines in Figure~\ref{fig:E_cuts} are fits to the cutoff energy $E_{\rm cut}$ and the index $s$, as a function of the characteristic distance between sources, using the functions 

\begin{equation}
    E_{\rm cut} = E_0 \exp\left[ \left( \frac{d_{\rm s}}{d_0}\right)^B-1\right], \ \ \ \ \ \ \ s =s_0\left( \frac{d_{\rm s}}{d_0}\right)^C,
    \label{eq:factor_H}
\end{equation}
 with $d_0 = 100$\,Mpc and the other parameters being listed in Table~\ref{table_H}  for each of the mass groups considered and accounting also for the dependence with the spectral index $\gamma$ (in the range between 1 and 3).  We note that for a given element the suppression at high energies is very similar whether the particle is a primary or a secondary from a heavier element, and the results reported apply to both cases.

\begin{table}[ht]
\centering
\begin{tabular}[t]{ *{6}{c} } 
\hline
   &  $E_0\ [{\rm EeV}]$ & B & $s_0$ & C \\ 
\hline
H & $ 140 - 13\gamma   $& $ -0.45$& $ 2.17 -0.14\gamma $ & $1.06$\\ 
\hline
He & $ 19.3  $& $ -0.21$ & $ 2.25 +0.09\gamma $ & $ 0.18  $\\ 
\hline
N & $ 42.2 - 0.8\gamma $ & $ -0.23  $& $ 2.60 +0.10\gamma $ & $0.16 $\\ 
\hline
Si & $76.5 - 5.8\gamma $ & $ -0.24$ & $ 3.68 -0.33\gamma $ & $ 0.17  $\\ 
\hline
Fe & $118 - 8\gamma $ & $ - 0.23  $ & $ 1.90 +0.14\gamma $ & $ 0.34 $\\ 
\hline
\end{tabular}
\label{table_H}
\caption{Parameters describing the high-energy suppression of the spectrum using eq.~(\ref{eq:factor_H}) for each element.}
\end{table}

For all mass groups,  the propagation is quasirectilinear near the cutoff energies and hence the average redshift of production of the particles observed  is very small at those energies, as can be seen in the right panel of Figure~\ref{fig:Redshift_5_groups}. 
This also implies that the attenuation at high energies is not sensitive to the source evolution adopted.
Thus, there should be essentially no difference between the high-energy suppressions for the NE and SFR scenarios, as can indeed be seen in the left panel of Figure~\ref{fig:SFR_NES}.

We have considered up to now the spectral suppression appearing below a few hundred EeV  appearing  due to photopion production for protons or to photodisintegration for nuclei. 
If the power-law spectrum $ \propto E^{-\gamma}$ were to extend up to much larger energies, the flux of particles is however expected to show a recovery from the pronounced exponential drop. This is due to the fact that the corresponding cross sections have a sharp increase above the threshold for which the associated processes become allowed, which leads to the strong suppressions, but then saturate at higher energies. Hence, essentially a constant fraction of the flux from the nearest sources would be expected to reach the observer at energies well beyond the previously discussed cut-off energies.

\begin{figure}[t] 
    \centering
    \includegraphics[width=0.5\textwidth]{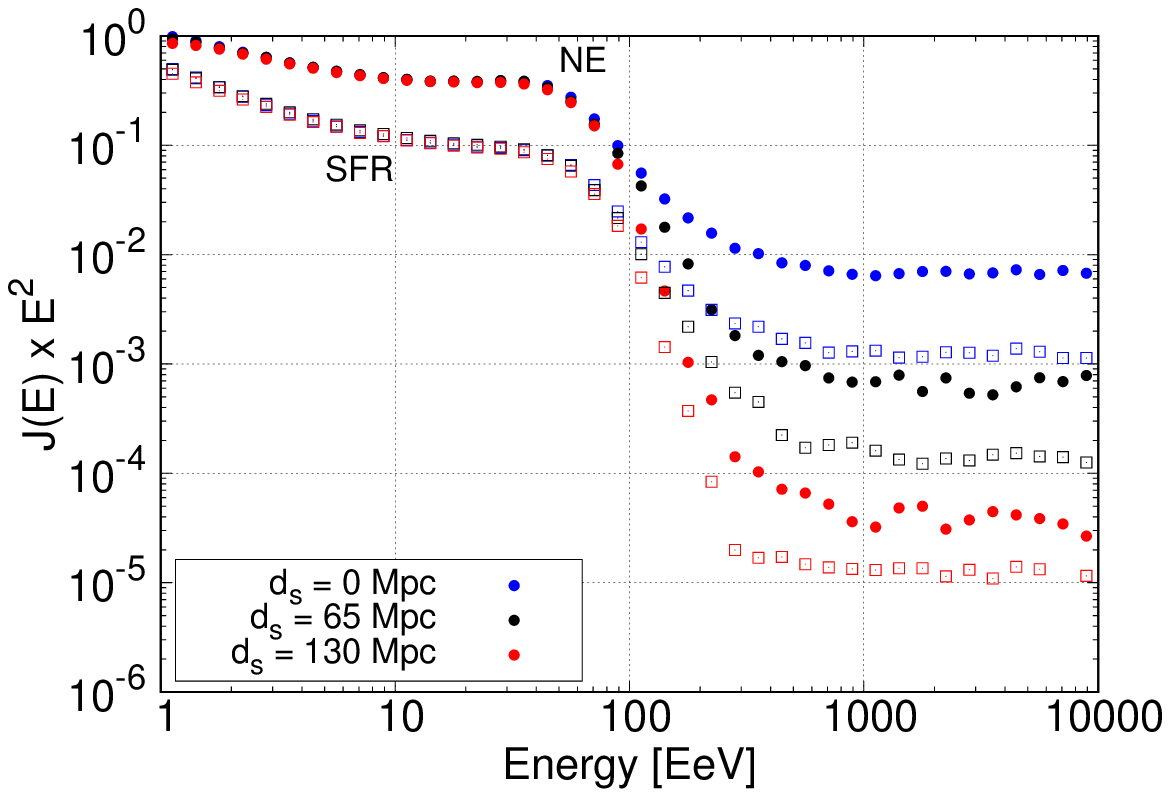}\includegraphics[width=0.5\textwidth]{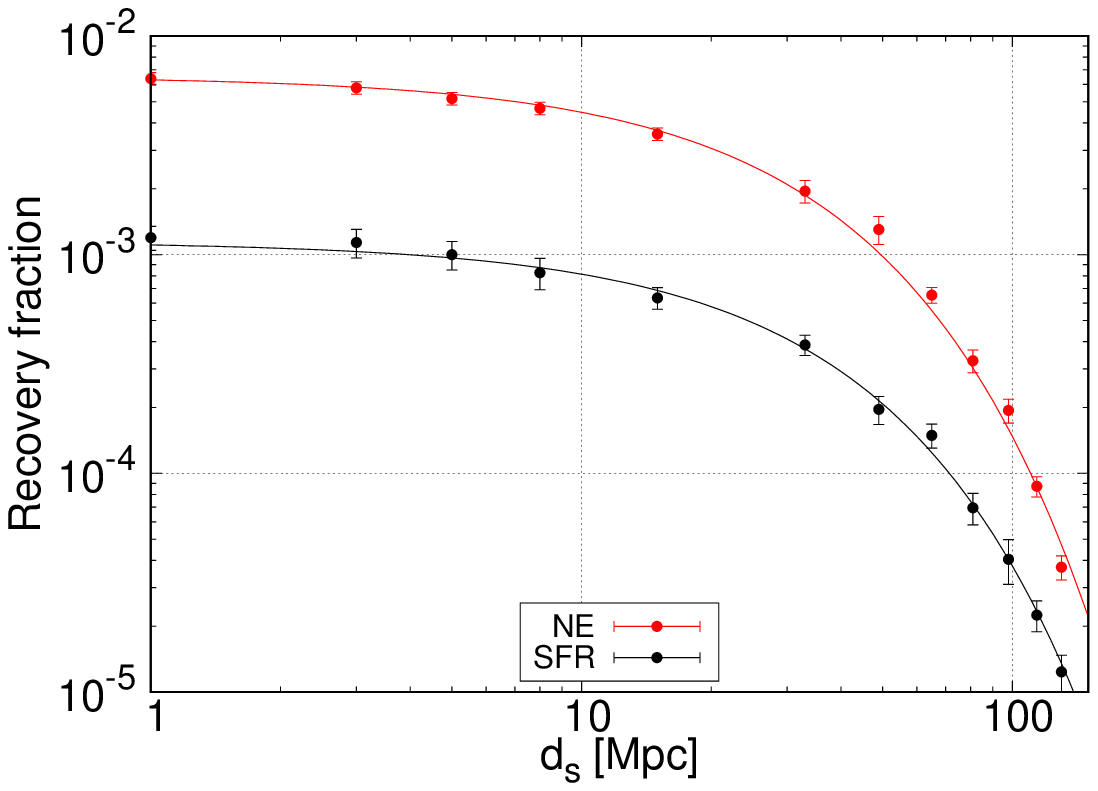}
    
    \caption{ Left panel: Flux of protons as a function of the energy for different densities of sources, normalised such that it tends to unity at very low energies, emphasising the recovery at the highest energies. Filled-in points correspond to the NE scenario while empty ones correspond to the SFR scenario. Right panel: Recovery fraction as a function of the characteristic distance between sources.}
    \label{fig:recovery}
\end{figure}

We show in the left panel of Figure~\ref{fig:recovery} the  flux of protons obtained when considering a spectrum $\propto E^{-2}$ that extends up to $10^5$\,EeV, for a continuous distribution of sources and for several discrete source densities, both  for the NE (filled dots) and SFR (open dots) scenarios. For all the densities considered the flux shows a recovery for energies above $\sim 200$\,EeV. The asymptotic recovery of the flux, corresponding to the high-energy plateau in the left panel, decreases for increasing separation between sources. The recovery  is actually smaller for the case of the SFR evolution than for the NE case  because the fraction of the flux contributed by the small redshift (and close-by sources) is smaller in this case. We show in the right panel of Figure~\ref{fig:recovery} the recovery fraction, defined as the ratio between the actual flux and the one that would be observed in the absence of interactions, as a function of the source separation, for both  evolution scenarios.
The recovery fraction decays exponentially with the distance between sources, as depicted by the solid lines which correspond to the function $F \exp(-d_{\rm s}/\delta)$. For the NE case we find $F_{\rm NE} = 6.5 \times 10^{-3}$, while for the SFR case we get $F_{\rm SFR} = 1.1 \times 10^{-3}$. The associated decay length is practically the same in both cases, being $\delta_{\rm NE}=26$\,Mpc and $\delta_{\rm SFR}=29$\,Mpc. This is expected as the flux at the highest energies comes from small redshifts, where the source evolution is unimportant.

\begin{figure}[t] 
    \centering
    \includegraphics[width=0.5\textwidth]{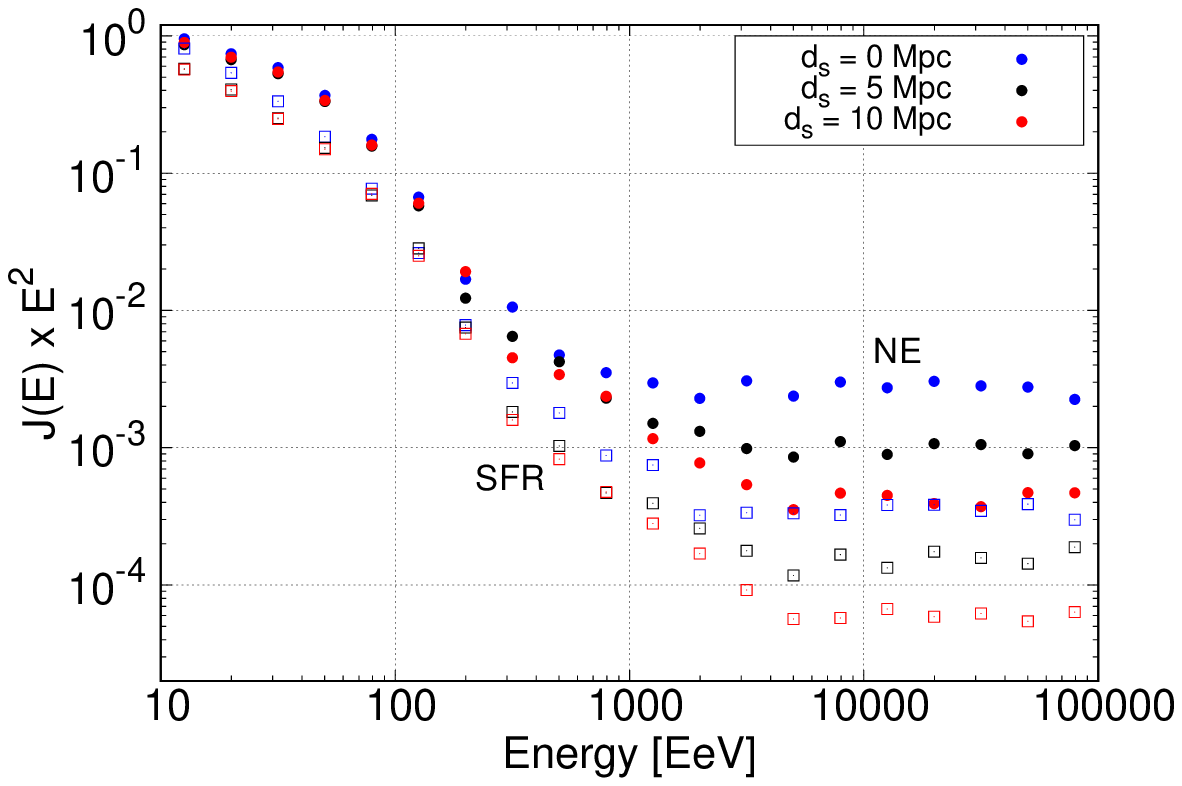}\includegraphics[width=0.5\textwidth]{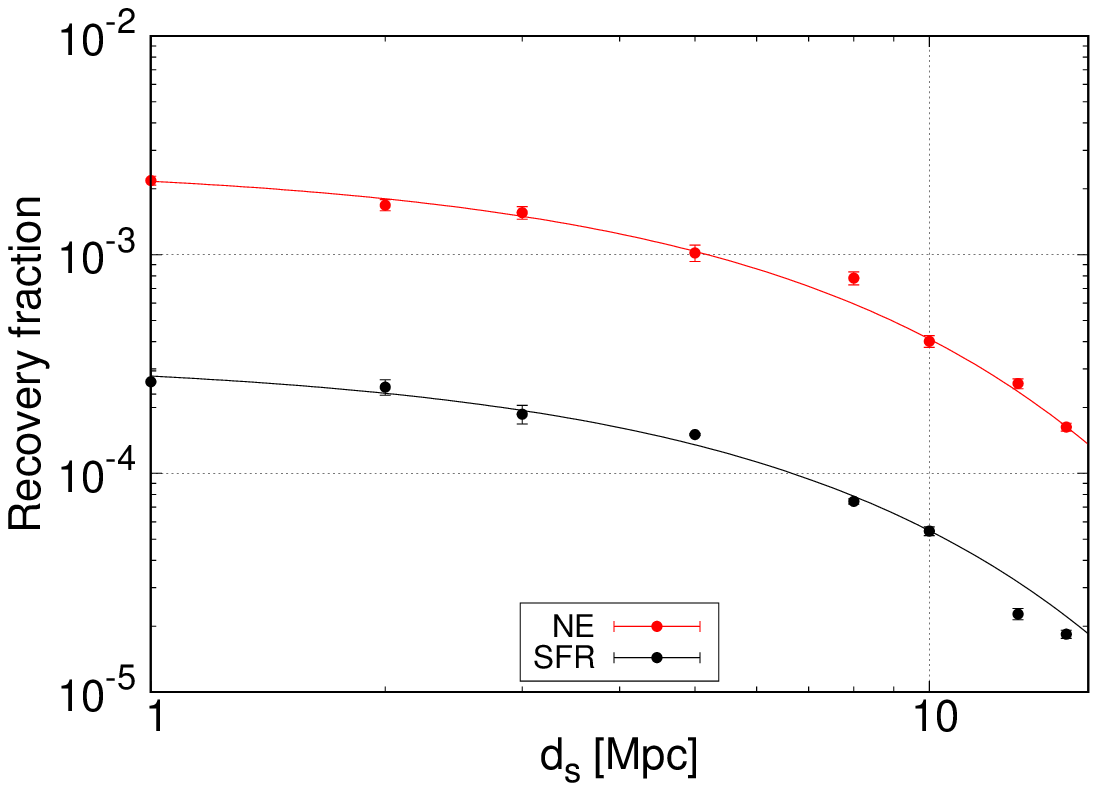}
    \caption{ Analogous of Figure~\ref{fig:recovery} but for an initial flux of Fe nuclei extending up to energies  of $10^5$\,EeV. The total flux from all secondaries is plotted.}
    \label{fig:recovery_Iron}
\end{figure}

Figure~\ref{fig:recovery_Iron} is similar to Figure~\ref{fig:recovery} but for an initial flux of Fe nuclei, extending up to $10^5$\,EeV. In this case the recovery fraction is smaller, given that the photodisintegration off CMB photons is very strong, leading to attenuation lengths of only a few Mpc. Indeed, the recovery fraction falls below $10^{-4}$ already for $d_s\simeq 10$\,Mpc. In this case the suppression can also be fitted using the exponential function $F \exp(-d_{\rm s}/\delta)$, where for the NE scenario we get $F_{\rm NE} =2.6 \times 10^{-3}$ while for the SFR case  $F_{\rm SFR} = 3.3 \times 10^{-4}$. The decay length is  very similar in both cases, being $\delta_{\rm NE}=5.5$\,Mpc and $\delta_{\rm SFR}=5.4$\,Mpc. Note that the decays length associated to the Fe and to the proton primaries are comparable to the corresponding associated   attenuation length  at these energies (see \cite{Progress_review} for a review), and this explains the differences in the corresponding  values obtained. One can see that the possibility of having a recovery of the flux is very strongly affected by the discreteness of the source distribution.

\section{DISCUSSION}

The flux of extragalactic cosmic rays reaching the observer from a discrete distribution of sources differs from that expected in the case of a  distribution that is continuous throughout space. On one side, a more pronounced suppression of the flux appears at the highest energies when the attenuation length of the particles due to the interactions with the radiation backgrounds becomes comparable to the separation between sources. In this case, even the flux from the closest sources will be significantly attenuated by the interactions. This can also strongly affect the potential recovery of the observed flux in case the spectrum at the sources were to continue up to energies well beyond this attenuation cutoff.
On the other hand, in the presence of a turbulent magnetic field the flux of particles reaching the observer from a discrete distribution of sources gets suppressed at low energies due to the magnetic horizon effect. This happens  when the diffusing  particles have not enough time to reach the observer even from the closest sources. We have quantified in detail both effects here, in such a way that the spectrum for the discrete source distribution can be obtained from the one in the continuous source distribution case through a multiplicative factor.

We have extended the magnetic suppression studies that were performed for the case of primary protons  in \cite{difu1} to the case of other primary nuclei accelerated at the sources as well as to the secondary nuclei resulting from photodisintegration during propagation. In all the cases the magnetic suppression depends on the magnetic-field parameters and the mean source separation  through the critical energy 
$E_{\rm c}= Z |e| B_{\rm rms} L_{\rm coh}$ and the parameter $X_{\rm s}\equiv d_{\rm s}/\sqrt{R_H L_{\rm coh}}$. We have shown that nuclei that reach the observer with a mass in the same mass group as the original one experience a magnetic suppression similar to that of primary protons, when taken as a function of $E/E_{\rm c}$. This is expected since particles with the same rigidity (and thus having the same value of $E/E_{\rm c}$) describe the same trajectories in the turbulent magnetic field.
We have also shown that secondary protons and the nuclei arriving with a mass significantly smaller than the primary ones have a milder magnetic suppression. Despite suffering photodisintegration during their trip, nuclei mostly travel with a constant rigidity, since the mass and the charge decrease in similar proportions, thus the trajectories of the secondary nuclei in the turbulent magnetic field are similar to those that their primaries would have had in the absence of photodisintegration.  The main reason explaining the milder suppression is that the nuclei arriving with a significantly smaller mass originated on average at higher redshifts, and thus have a longer available time to reach the observer from the closest sources. In the case of the secondary protons, there is also another effect entering into play because the rigidity of the particle is not constant during the trip since nuclei have approximately twice the rigidity of the secondary protons that they produce. Because of this, secondary protons can arrive from sources farther away, and their suppression is then even milder.

\begin{figure}[t]
    \centering
    \includegraphics[width=0.6\textwidth]{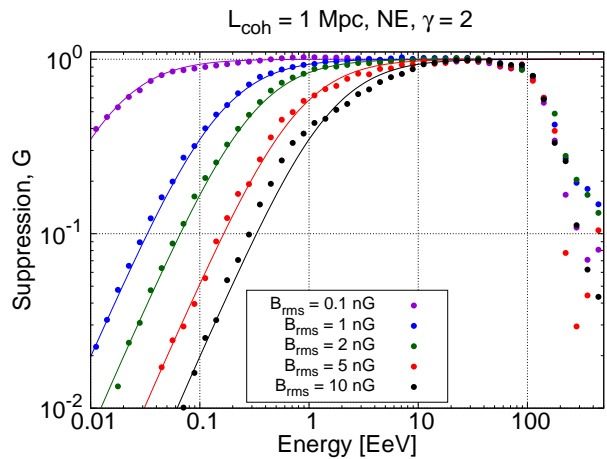}
    \caption{Suppression of the proton flux for different amplitudes of the turbulent magnetic field.}
    \label{fig:J_B}
\end{figure}

In the examples that we have shown, the magnetic suppression took place for energies lower than those at which pair production significantly affects the propagation of the particles. In this case, the effects on the flux reaching the observer due to the interactions with the radiation backgrounds and that due to the magnetic fields get factorised, and hence the multiplication of the flux in the absence of magnetic fields  times  the magnetic suppression factor $G(E/E_{\rm c})$ 
leads to a very good description of the flux from a discrete distribution of sources in the presence of a turbulent field. 
Since the energy $E_{0.5}$ at which the magnetic suppression becomes important, i.e. such that $G(E_{0.5}/E_{\rm c})=0.5$, is approximately $E_{0.5} \sim X_{\rm s} E_{\rm c}/5$, the factorisation holds if this energy is smaller than about 1\,EeV.
This corresponds to values of $X_{\rm s} E_{\rm c} \lesssim 5$\,EeV.  Figure~\ref{fig:J_B} shows  the suppression of the flux for a discrete distribution of sources with $X_{\rm s}=1$, $L_{\rm coh} = 1$\,Mpc and different values of the rms amplitude of the magnetic field (points). We see that for $B_{\rm rms}$ values larger than 5~nG, corresponding to $ E_{\rm c} \gtrsim 5$\,EeV, a slight departure from the analytic expression  $G(E/E_{\rm c})$, depicted by the solid lines, appears, as was discussed in ref.~\cite{difu1}. The difference comes from the fact that pair production leads to a decrease in the rigidity of the particles as they propagate and hence  an interplay between the attenuation and magnetic horizon effects appears. In the case of nuclei, although the pair production cross section grows as $Z^2$, the associated inelasticities decrease as $1/A$ and as a result the associated attenuation length for pair production losses becomes comparable to the one of adiabatic losses at higher energies (around 50\,EeV for Fe nuclei), and hence the change in the magnetic suppression shape due to pair production would happen in this case for larger values of $X_{\rm s}E_{\rm c}$ than for the proton case.

Let us note that the composition and spectrum observations from the Pierre Auger Observatory above the ankle require the presence of a mixed composition in which increasingly heavier elements become dominant as the energy increases, with little overlap among them. To suppress this overlap, it is necessary to have a strong suppression of the heavy elements for decreasing energies. This can result from elementary spectra with a very hard spectral index at the sources ($\gamma<1$), which is however at odds with expectations from second order Fermi acceleration, in combination with a relatively low rigidity spectral cutoff to suppress the light component at high energies \cite{combinedfit}.
Alternatively, the magnetic suppression effect discussed in the present work, with the associated  hardening of the spectrum for low rigidities, has been proposed as a possible explanation for the composition and spectrum observations \cite{difu1}.
 The suppression of the spectrum at low energies resulting from the magnetic horizon discussed here can instead make the effectively hard spectrum reaching the observer to become compatible with a spectral index at the source closer to two \cite{difu1,icrcwitt,mr20}. The results obtained in this work should allow to obtain refined predictions for this effect in different scenarios. 

The presence of an extragalactic magnetic field also affects the distribution of the CR arrival directions. For each individual source, one expects a transition from a pointlike image at high rigidity to an increasingly spread distribution, tending to a dipole, for small rigidity, as studied in \cite{hmr16}. The total arrival direction distribution as a function of the energy resulting from an ensemble of sources will depend on the distance and direction of the sources, the mass composition of the particles and the magnetic field parameters. In particular, the evolution with energy of the dipolar component of the distribution in a realistic scenario has been obtained in \cite{HMR2015}, showing a good agreement with the results from the Pierre Auger Observatory \cite{ApJ2018}. As discussed in \cite{HMR2015}, the dipole amplitude has only a mild dependence on the turbulent extragalactic magnetic field strength, due to a cancellation between the diffusive enhancement of the contribution to the CR density from the nearby sources and the simultaneous reduction of their dipolar amplitudes.

\section*{Acknowledgments}
This work was supported by CONICET (PIP 2015-0369) and ANPCyT (PICT 2016-0660). We are very grateful to D. Boncioli for help with SimProp and useful discussions, and to the authors of SimProp for making it available. We also thank D. Harari for useful discussions.

\end{document}